\documentclass[aps,pre,showpacs,twocolumn]{revtex4}
\usepackage{bm}
\usepackage{graphicx}
\bibstyle{approve.bib}
\usepackage{amssymb}
\usepackage{amsmath}
\usepackage{esint}
\usepackage{epstopdf}
\usepackage{color}
\usepackage{gensymb}
\usepackage{mathtools}
\usepackage{suffix}

\newcommand{\be}{\begin{equation}}
\newcommand{\ee}{\end{equation}}
\newcommand{\bea}{\begin{eqnarray}}
\newcommand{\eea}{\end{eqnarray}}

\newcommand{\br}{\mathbf{r}}

\newcommand{\bk}{\mathbf{k}}

\newcommand{\e}{\varepsilon}

\newcommand{\tv}{\tilde{v}}
\newcommand{\tdel}{\tilde{\Delta}}

\newcommand{\tz}{\tilde{z}_a}
\newcommand{\tL}{\tilde{L}}
\newcommand{\tzt}{\tilde{z}_t}

\newcommand{\tbr}{\tilde{\br}}

\newcommand{\bz}{\bar{z}_a}

\newcommand{\bL}{\bar{L}}
\newcommand{\pa}{\parallel}

\DeclarePairedDelimiterX\MeijerM[3]{\lparen}{\rparen}
{\begin{smallmatrix}#1 \\ #2\end{smallmatrix}\delimsize\vert\,#3}

\newcommand\MeijerG[8][]{G^{\,#2,#3}_{#4,#5}\MeijerM[#1]{#6}{#7}{#8}}

\WithSuffix\newcommand\MeijerG*[7]{G^{\,#1,#2}_{#3,#4}\MeijerM*{#5}{#6}{#7}}

\begin{document}

\title{Correlation-induced DNA adsorption on like-charged membranes}

\author{Sahin Buyukdagli$^{1}$\footnote{email:~\texttt{Buyukdagli@fen.bilkent.edu.tr}}  
and Ralf Blossey$^{2}$\footnote{email:~\texttt{ralf.blossey@iri.univ-lille1.fr}}}
\address{$^1$Department of Physics, Bilkent University, Ankara 06800, Turkey\\
$^2$Univ. Lille 1, CNRS, UGSF UMR8576, 59000 Lille, France}

\date{\small\it \today}

\begin{abstract}
The adsorption of DNA or other polyelectrolyte molecules on charged membranes is a recurrent motif in soft matter and bionanotechnological systems. Two typical situations encountered are the deposition of single DNA chains onto substrates for further analysis, e.g. by force microscopy, or the pulling of polyelectrolytes into membrane nanopores, as in sequencing applications. In this paper we present a theoretical analysis of such scenarios based on the self-consistent field theory approach, which allows us to address the important effect of charge correlations. We calculate the grand potential of a \textcolor{black}{stiff} polyelectrolyte immersed in an electrolyte in contact with a negatively charged dielectric membrane. \textcolor{black}{For the sake of conciseness, we neglect conformational polymer fluctuations and model the molecule as a rigid charged line.} At strongly charged membranes, the adsorbed counterions enhance the screening ability of the interfacial region. In the presence of highly charged polymers such as double-stranded DNA molecules close to the membrane, this enhanced interfacial screening dominates the mean-field level DNA-membrane repulsion and results in the adsorption of the DNA molecule to the surface. \textcolor{black}{This picture provides a simple explanation for the recently observed DNA binding onto similarly charged substrates [G. L.-Caballero et al., Soft Matter \textbf{10}, 2805 (2014)] and points out charge correlations as a non-negligible ingredient of polymer-surface interactions.}
\end{abstract}


\date{\today}
\maketitle

\section{Introduction}

Modern bionanotechnology advances at an ever increasing speed, with theoretical understanding trailing sometimes behind. In several instances, this is due to the fact that for a theoretical understanding of many relevant applications the mathematical basis is still not entirely laid. In this paper, we address one such case: the adsorption of polyelectrolytes such as DNA molecules on like-charged membranes. This is a practical problem for at least two situations: i) the adsorption of charged molecules on substrates for further analysis by, e.g., force microscopy \cite{lee12}, and ii) the approach of polyelectrolytes to membrane nanopores in sequencing applications \cite{zwolak08}.  Although these processes have been previously modeled within mean-field (MF) electrostatics~\cite{pod1,pod2,dun,orland,muthu}, in the theoretical analysis of such situations, the difficulty lies in the inclusion of charge correlations at like-charged membranes. Indeed, the distortion of the ionic environment by the membrane charges requires a theoretical treatment beyond the classical MF-level 
Poisson-Boltzmann equation. Rigorous methods to treat such effects have emerged a few years ago and are only recently beginning to be applied and tested in relevant physical situations.  Without such a theoretical framework, the physical treatment has to rely on ad-hoc approaches based on uncontrolled approximations and also difficult to generalize. With the idea of an immediate application in mind, more microscopic theoretical approaches such as atomistic simulations often try to cover many specific details of charge liquids. The resulting complexity leads to a lack of analytical understanding of the main underlying effects, see e.g. the recent review~\cite{faraudo13}. 

Within this logic in mind, we consider the polyelectrolyte adsorption problem from a purely electrostatic perspective. Let us take as a specific starting point the work by
Sens and Joanny who, in Ref.~\cite{Sens2000}, determined the self-energy of a stiff polymer of line charge $-\tau$ in the vicinity of a charged wall with surface charge density 
$\sigma$. In terms of the Gouy-Chapman length $\mu=1/(2\pi \ell_B \sigma)$ with $\ell_B$ the Bjerrum length, they determined the asymptotic behaviour of the free energy as 
a function of the height $h$ of the  polymer above the surface:
\begin{equation}
\delta {\cal F}(h) \simeq \frac{1}{2}\ell_B \tau^2 \left[\frac{2}{3} + \ln\left(\frac{4\pi}{3} \frac{h}{a} \right) \right]
\end{equation}
for $h \gg \mu $ and
\begin{equation}
\delta {\cal F}(h) \simeq \frac{1}{2}\ell_B \tau^2 \left[-\Gamma + \frac{2\pi}{3\sqrt{3}} + \ln\left(\frac{\pi \mu^2}{ah} \right) \right]
\end{equation}
for the opposite limit, $h \ll \mu$. In both equations, $a$ is a cutoff, taken as the diameter of the DNA molecule.

Their  result is obtained from an approximate calculation employing the solution of the Poisson-Boltzmann equation $\phi(z)$ reduced to the Gouy-Chapmann limit of low salt, considering the fluctuations $\delta \phi(z)$ around this solution, and taking the boundary condition at the polymer into account. The free energy is then calculated from a charging process, according to the expression
\begin{equation} \label{fe}
\delta{\cal F}(h) = \int_0^{\tau} d\tau' \left[\phi(z) + \delta \phi(z) \right]_{z=h},
\end{equation}
where the first term is the MF-level interaction free energy between the polymer and the wall. The chief interest lies in the determination of the second fluctuation term associated with charge fluctuations mediated by the mobile ions of the solution.

\begin{figure}
\includegraphics[width=1\linewidth]{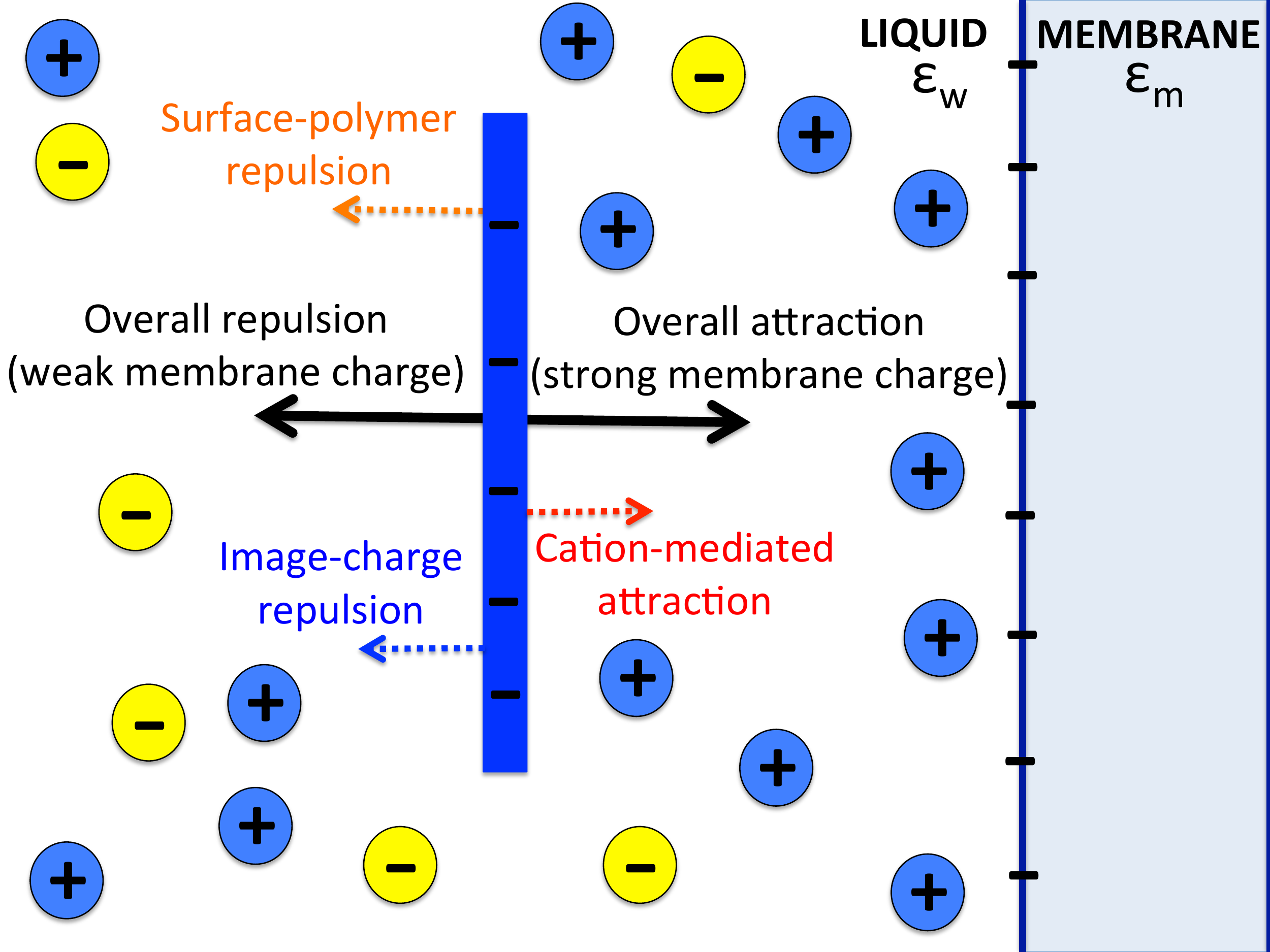}
\caption{(Color online) Schematic representation of the negatively charged polymer immersed in a symmetric electrolyte and interacting with the like-charged dielectric membrane. Electrostatic forces acting on the polymer (dashed arrows) : MF-level membrane-polymer repulsion (orange),  beyond-MF repulsive image charge force (blue) and attractive cation-mediated solvation force (red).}
\label{fig0}
\end{figure}

\textcolor{black}{In this paper, we calculate the energetic cost to drive a stiff polyelectrolyte immersed in a charged liquid from the bulk region to the neighborhood of a  like-charged dielectric membrane. Our calculation generalizes the Debye-H\"{u}ckel (DH) theories of polymer-membrane interactions~\cite{Netz2000,Buyuk2016} to strongly charged membranes.  The system, composed of the negatively charged polymer, the like-charged dielectric membrane, and the electrolyte is depicted in Fig.~\ref{fig0}.  For the sake of simplicity, we neglect conformational polymer fluctuations and model the polyelectrolyte as a rigid line charge. In the calculation of the work required to drive the polymer to the membrane surface, our starting point is the variational grand potential of the charged system.  First, we expand this potential in the electrostatic coupling parameter and keep the leading term of this expansion. In order to reduce the grand potential to an analytically tractable form, we perform a second expansion in terms of the polymer charge density. This derivation is presented in Appendix~\ref{ap1} in detail and its result summarised in Sec.~\ref{forpot}. The resulting polymer grand potential is a generalisation of Eq.~(\ref{fe}) within the self-consistent field theory. In Section II B, we calculate the one-loop level electrostatic Green's function required for the explicit evaluation of the polymer grand potential in plane geometry.} 

\textcolor{black}{Within this formalism, Sections III and IV focus, respectively, on the polymer adsorption and the polymer approach prior to translocation events. Therein, we characterize quantitatively electrostatic many-body effects on polymer-membrane interactions.  We find that the nature of these interactions is determined by the competition between three electrostatic force components illustrated in Fig.~\ref{fig0}. The direct polymer-membrane charge coupling results in the standard MF like-charge repulsion (orange arrow). This repulsive force is enhanced by the polymer-image charge interactions (blue arrow) induced by the dielectric contrast between the low permittivity membrane and the solvent. The third contribution is due to the cation attraction by the charged membrane. The resulting counterion excess increases the screening ability of the interfacial region. This means a lower interfacial polymer free energy with respect to bulk, which translates into an attractive force oriented to the membrane surface (red arrow). At strongly charged membranes, the cation-induced attractive force dominates the repulsive components, resulting in the like-charge adsorption of the polymer onto the membrane. This result provides a simple explanation for the recent experimental observation of this peculiarity~\cite{Molina2014}. The limitations and possible extensions of the present theory are discussed in the Conclusion part.}

\section{Formulation of the problem}

\subsection{Polymer grand potential}
\label{forpot}

The electrostatic grand potential of a stiff polymer with charge density $\sigma_p(\br)$ immersed in a charged liquid in contact with a solid membrane is given by
\be\label{Eq0}
\Omega_p=\Omega_{pm}+\Omega_{pp},
\ee
where the MF-level polymer-membrane charge interaction is
\be
\label{Eq0II}
\Omega_{pm}=\int\mathrm{d}\br\sigma_p(\br)\psi_{0m}(\br),
\ee
and the polymer self-energy including charge correlations reads
\be
\label{Eq0I}
\Omega_{pp}=\frac{1}{2}\int\mathrm{d}\br\mathrm{d}\br'\sigma_p(\br)v(\br,\br')\sigma_p(\br').
\ee
In Eq.~(\ref{Eq0II}), the electrostatic potential $\psi_{0m}(\br)$ induced by the membrane charge $\sigma_m(\br)$ solves the PB equation
\bea\label{EqII}
&&\nabla\e(\br)\cdot\nabla\psi_{0m}(\br)-\frac{2\rho_bqe^2}{k_BT}e^{-V_i(\br)}\sinh\left[q\psi_{0m}(\br)\right]\nonumber\\
&&=-\frac{e^2}{k_BT}\sigma_m(\br).
\eea
Furthermore, the one loop-level (1l) electrostatic Green's function $v(\br,\br')$ of Eq.~(\ref{Eq0I}) is the solution of the kernel equation
\bea
\label{EqIII}
&&\nabla\e(\br)\cdot\nabla v(\br,\br')-\frac{2\rho_bq^2e^2}{k_BT}e^{-V_i(\br )}\cosh\left[q\psi_{0m}(\br)\right]v(\br,\br')\nonumber\\
&&=-\frac{e^2}{k_BT}\delta(\br-\br').
\eea
The derivation of these expressions is presented in detail in Appendix~\ref{ap1}.

\begin{figure}
\includegraphics[width=1\linewidth]{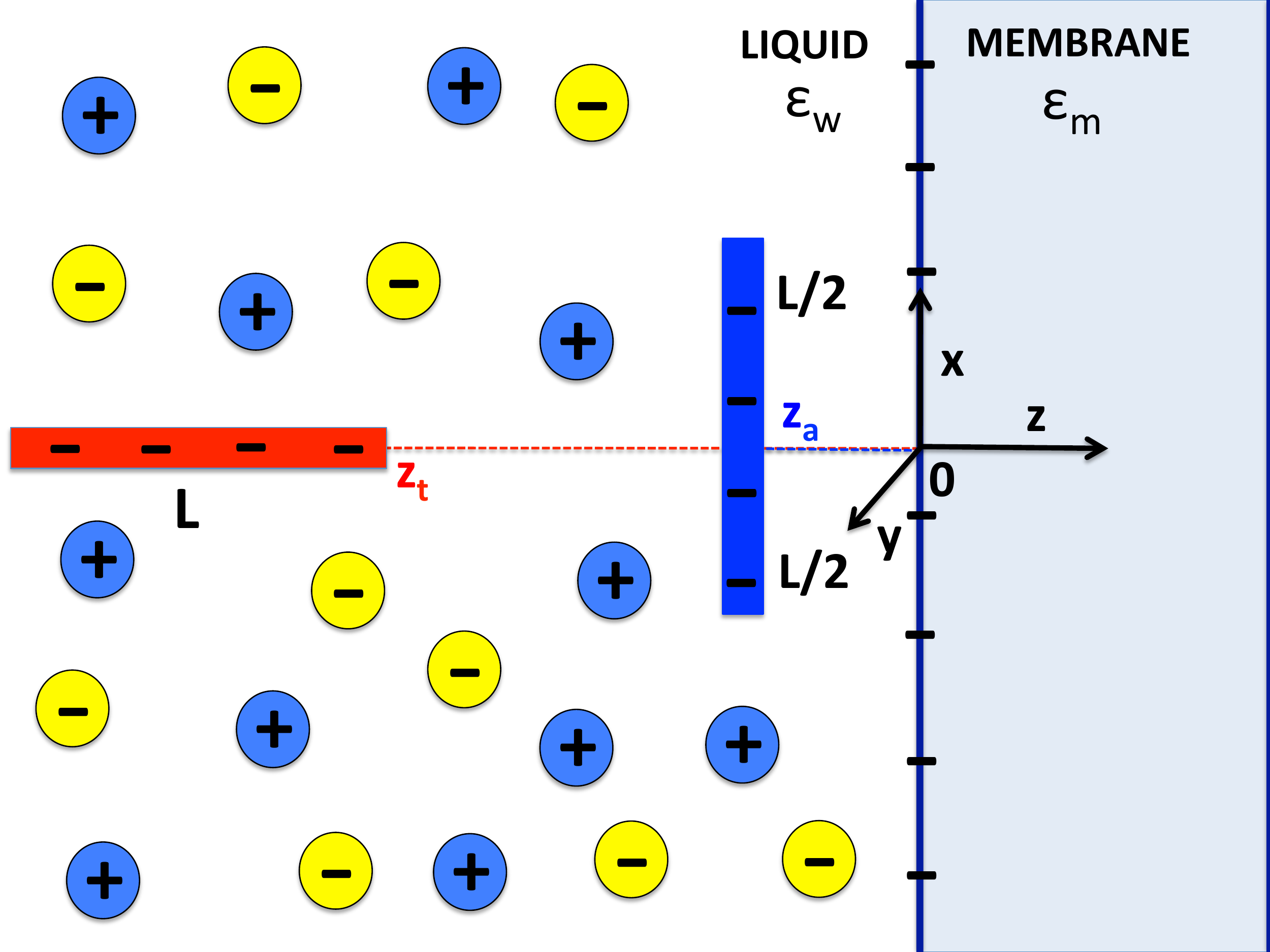}
\caption{(Color online) Negatively charged polyelectrolyte of length $L$ and linear charge density $\tau$, with the total charge $Q_p=L\tau$. The right end of the approaching polymer (red) is located at $z=z_t<0$ and the adsorbing polymer (blue) at $z=z_a<0$ from the membrane surface located at $z=0$.}
\label{fig1}
\end{figure}

In Eqs.~(\ref{EqII})-(\ref{EqIII}), the function $\e(\br)$ is the dielectric permittivity profile, $\rho_b$ the bulk salt density, $q$ the ion valency, $e$ the electron charge, $k_B$ the Boltzmann constant, and $T=300$ K stands for the ambient temperature. Moreover, the steric potential $V_i(\br)$ excludes the mobile ions from the location of hard bodies such as the membrane matrix (see Fig.~\ref{fig1}). Finally, we note that in the present work, energies will be expressed in units of the thermal energy $k_BT$.

The physically relevant parameter is the work to be done adiabatically in order to bring the polymer from the bulk region to the neighbourhood of the membrane. This corresponds to the difference between the grand potential~(\ref{Eq0}) and the bulk grand potential,
\be
\label{np}
\Delta\Omega_p=\Delta\Omega_{pp}+\Omega_{pm},
\ee
where the renormalised self-energy follows from Eq.~(\ref{Eq0I}) as
\be
\label{np2}
\Delta\Omega_{pp}=\frac{1}{2}\int\mathrm{d}\br\mathrm{d}\br'\sigma_p(\br)\left[v(\br,\br')-v_b(\br-\br')\right]\sigma_p(\br').
\ee
In Eq.~(\ref{np2}), we introduced the bulk solution of Eq.~(\ref{EqIII}) corresponding to the spherically symmetrical DH Green's function
\be
\label{dhb}
v_b(\br-\br')=l_B\frac{e^{-\kappa_b|\br-\br'|}}{|\br-\br'|},
\ee
with the DH screening parameter $\kappa_b^2=8\pi\ell_Bq^2\rho_b$ and the Bjerrum length $\ell_B=e^2/(4\pi\e_wk_BT)$, where $\e_w=80$ stands for the dielectric permittivity of water. Infinitely far from the membrane where the external potential $\psi_{0m}(\br)$ in Eq.~(\ref{Eq0II}) is zero and the Green's function $v(\br,\br')$ tends to the bulk limit~(\ref{dhb}), the net grand potential~(\ref{np}) vanishes.

\subsection{Solution to the PB and electrostatic kernel equations}

\label{ker}
We calculate here the MF potential $\psi_{0m}(\br)$ and the Green's function $v(\br,\br')$ required for the evaluation of the polymer grand potential~(\ref{np}). In the case of a planar dielectric interface located at $z=0$ and carrying a negative surface charge $\sigma_m(\br)=-\sigma_m\delta(z)$, with the electrolyte located on the left half-space $z<0$ (see Fig.~\ref{fig1}), the MF potential is given by~\cite{Isr}
\be\label{EqIV}
\psi_{0m}(z)=-\frac{2}{q}\ln\left[\frac{1+e^{\kappa_b(z-z_0)}}{1-e^{\kappa_b(z-z_0)}}\right],
\ee
where we introduced the characteristic thickness of the interfacial counterion layer $z_0=-\ln(\gamma_c(s))/\kappa_b$~\cite{Lau2008}, the auxiliary parameter 
\be\label{aux}
\gamma_c(s)=\sqrt{s^2+1}-s, 
\ee
the dimensionless parameter $s=\kappa_b\mu$, and the Gouy-Chapman length $\mu=1/(2\pi q\ell_B\sigma_m)$. 

Accounting for the plane geometry, one can Fourier-expand the Green's function as
\be\label{fourgr}
v(\br,\br')=\int\frac{d^2\bk}{4\pi^2}e^{i\bk\cdot\br_{\pa}}\tv(z,z';k),
\ee
where the vector $\br_{\pa}$ indicates any point located in the $x-y$ plane that coincides with the membrane wall. Injecting the expansion~(\ref{fourgr}) together with the membrane potential~(\ref{EqIV}) into Eq.~(\ref{EqIII}), the one-loop level kernel equation follows as
\bea\label{forker}
&&\frac{\partial}{\partial z}\e(z)\frac{\partial}{\partial z}\tv(z,z';k)\textcolor{black}{-\e_m\theta(z)k^2\tv(z,z';k)}\nonumber\\
&&-\textcolor{black}{\e_w}\theta(-z)\left\{p^2+2\kappa_b^2\mathrm{csch}^2\left[\kappa_b(z-z_0)\right]\right\}\tv(z,z';k)\nonumber\\
&&=-\frac{e^2}{k_BT}\delta(z-z').
\eea
In Eq.~(\ref{forker}), we introduced the dielectric permittivity profile
\be\label{eqdi}
\e(z)=\e_w\theta(-z)+\e_m\theta(z)
\ee
with $\e_m$ the membrane permittivity, the screening function $p=\sqrt{k^2+\kappa_b^2}$, and the Heaviside step function restricting the location of the mobile charges to the half-space located at $z<0$. We need the solution of Eq.~(\ref{forker}) for the source charge located in the left half-space, i.e. $z'<0$.  In this case, the general solution is given by
\bea\label{gr1}
\tv(z,z',k)&=&c_1h_-(z)\theta(-z)\theta(z'-z)\\
&&+\left[c_2h_-(z)+c_3h_+(z)\right]\theta(-z)\theta(z-z')\nonumber\\
&&+c_4e^{-kz}\theta(z), \nonumber
\eea
where the increasing and decreasing homogeneous solutions of Eq.~(\ref{forker}) read for $z<0$
\be\label{solpar}
h_\pm(z)=e^{\mp pz}\left\{1\pm\frac{\kappa_b}{p}\coth\left[\kappa_b(z-z_0)\right]\right\}.
\ee
The final step consists in calculating the integration constants $c_{1...4}$ in Eq.~(\ref{gr1}) by imposing the boundary conditions to be satisfied by the Green's function and the displacement field at $z=0$ and $z=z'$,
\bea\label{boun1}
&&\tv\left(z=0_-\right)=\tv\left(z=0_+\right)\\
\label{boun2}
&&\tv\left(z=z'_-\right)=\tv\left(z=z'_+\right)\\
\label{boun3}
&&\e(z)\left.\frac{\partial\tv}{\partial z}\right|_{z=0_-}=\e(z)\left.\frac{\partial \tv} {\partial z}\right|_{z=0_+}\\
\label{boun4}
&&\left.\frac{\partial\tv}{\partial z}\right|_{z=z'_+}-\left.\frac{ \partial \tv}{\partial z}\right|_{z=z'_-}=-4\pi\ell_B.
\eea
One finally gets for $z,z'<0$ the Fourier-transformed Green's function as
\be\label{tv1}
\tv(z,z';k)=\frac{2\pi\ell_B p}{k^2}\left[h_+(z_>)+\Delta(k) h_-(z_>)\right]h_-(z_<),
\ee
with $z_<=\mathrm{min}(z,z')$, $z_>=\mathrm{max}(z,z')$, and the auxiliary function
\be\label{del}
\Delta(k)=\frac{\kappa_b^2\mathrm{csch}^2\left(\kappa_bz_0\right)+(p-\eta k)\left[p-\kappa_b\coth\left(\kappa_bz_0\right)\right]}
{\kappa_b^2\mathrm{csch}^2\left(\kappa_bz_0\right)+(p+\eta k)\left[p+\kappa_b\coth\left(\kappa_bz_0\right)\right]},
\ee
where we introduced the parameter $\eta=\e_m/\e_w$ accounting for the dielectric discontinuity at $z=0$.

\section{Polymer adsorbing to the membrane}
\label{ad}

We now consider the polymer binding to the membrane surface where the polymer orientation is parallel with the membrane wall (see Fig.~\ref{fig1}). Because the membrane wall is assumed to be infinitely large, the polymer grand potential will depend only on the distance from the membrane and the location of the origin of the coordinate system is irrelevant. Thus, for the sake of simplicity, we choose the Cartesian coordinate system with the upper half of the polymer  oriented along the positive $x$-axis and the lower half along the negative $x$-axis. Furthermore, we place the centre of mass of the polymer at $y=0$ and at a distance $z_a$ from the membrane. In this configuration, the polymer charge density reads
\be
\label{polch}
\sigma_p(\br)=-\tau\;\theta(x+L/2)\;\theta(L/2-x)\delta(y)\delta(z-z_a).
\ee

Injecting the density function~(\ref{polch}) into Eq.~(\ref{Eq0II}) together with the membrane potential~(\ref{EqIV}), the MF-level polymer-membrane interaction potential takes the form
\be
\label{pm1}
\Omega_{pm}(z_a)=\frac{2L\tau}{q}\ln\left[\frac{1+e^{\kappa_b(z_a-z_0)}}{1-e^{\kappa_b(z_a-z_0)}}\right].
\ee
In order to obtain the polymer self-energy, we insert first the Fourier expansion of the Green's function~(\ref{fourgr}) and the charge density function~(\ref{polch}) into Eq.~(\ref{np2}). This yields
\bea
\label{pp1}
\Delta\Omega_{pp}(z_a)&=&\frac{\tau^2}{2}\int\frac{\mathrm{d}^2\bk}{4\pi^2}\;\left[\tv(z_a,z_a;k)-\tv_b(0;k)\right]\\
&&\times\int_{-L/2}^{L/2}\mathrm{d}x_1\int_{-L/2}^{L/2}\mathrm{d}x_2\;e^{i\bk\cdot(\mathbf{x}_1-\mathbf{x}_2)},\nonumber
\eea
where we introduced the Fourier transformed bulk Green's function
\be
\tv_b(z-z')=\frac{2\pi\ell_B}{p}e^{-p|z-z'|}.
\ee
In Eq.~(\ref{pp1}), the integrals over $x_1$ and $x_2$ correspond to transverse charge correlations parallel with the wall. Evaluating the spatial integrals and substituting into Eq.~(\ref{pp1}) the Green's function~(\ref{tv1}), after some algebra, one gets the rescaled self-energy in the form
\bea
\label{pp3}
\Delta\Omega_{pp}(z_a)&=&\frac{L\ell_B\tau^2\kappa_b^2}{2}\iint_{-\infty}^{+\infty}\frac{\mathrm{d}k_x\mathrm{d}k_y}{k^2p}\frac{2\sin^2\left(k_xL/2\right)}{\pi Lk_x^2}\nonumber\\
&&\times\left\{-\mathrm{csch}^2\left[\kappa_b(z_a-z_0)\right]\right.\nonumber\\
&&\hspace{0cm}\left.+\Delta(k)\left(\frac{p}{\kappa_b}-\mathrm{coth}\left[\kappa_b(z_a-z_0)\right]\right)^2e^{2pz_a}\right\}.\nonumber\\
\eea
According to Eq.~(\ref{np}), the net grand potential of the polymer located at the distance $z_a$ from the membrane is given by the superposition of Eqs~(\ref{pm1}) and~(\ref{pp3}), 
\be\label{ptot1}
\Delta\Omega_p(z_a)=\Delta\Omega_{pp}(z_a)+\Omega_{pm}(z_a).
\ee

\subsection{Counterion liquids}
\label{countlim}

We consider herein the simplest case of a charged liquid exclusively composed of counterions. In order to evaluate the polymer grand potential in this regime, we take the vanishing salt limit $\kappa_b\to0$ of Eqs.~(\ref{pm1}) and~(\ref{pp3}) in which case the polymer-self-energy takes the form
\bea\label{pp4}
\Delta\Omega_{pp}(z_a)&=&\frac{L\ell_B\tau^2}{2\left(\mu-z_a\right)^2}\iint_{-\infty}^{+\infty}\frac{\mathrm{d}k_x\mathrm{d}k_y}{k^3}\frac{2\sin^2\left(k_xL/2\right)}{\pi Lk_x^2}\nonumber\\
&&\hspace{0mm}\times\left\{-1+\Delta_c(\mu k)\left[1+(\mu-z_a)k\right]^2e^{2kz_a}\right\}.\nonumber\\
\eea
\textcolor{black}{In the same counterion-only limit}, the polymer-membrane charge coupling energy reads~\cite{prc} 
\be
\label{pm2}
\Omega_{pm}(z_a)=-\frac{2L\tau}{q}\ln(1-z_a/\mu).
\ee
In Eq.~(\ref{pp4}), we introduced the \textcolor{black}{vanishing salt} limit of the function $ \Delta$ given by Eq.~(\ref{del}),
\be\label{delc}
\Delta_c(x)=\frac{1+(1-\eta)x(x-1)}{1+(1+\eta)x(x+1)}.
\ee 
One could simplify the double integral in Eq.~(\ref{pp4}) by considering the long polymer limit $L\to\infty$ where the sinusoidal function becomes a Dirac delta distribution, 
\be\label{dir}
\lim_{L\to\infty}\frac{2\sin^2\left(k_xL/2\right)}{\pi Lk_x^2}=\delta(k_x),
\ee
removing the integral on the wavevector $k_x$. In this limit, setting $\e_m=0$ (or $\eta=0$), Eq.~(\ref{pp4}) tends to the polymer self-energy calculated in Ref.~\cite{Sens2000} via the charging procedure. We should however note that in the present counterion-only regime, the limit $L\to\infty$ results in the infrared (IR) divergence of the self-energy.  

For the sake of analytical transparency, we will focus on the opposite limit of short polymers. Within this approximation, one can Taylor-expand the sinusoidal function in Eq.~(\ref{pp4}) at the leading order in the polymer length $L$ and recover the plane symmetry in the reciprocal $(k_x,k_y)$ plane. At the next step, we transform to polar coordinates ($k,\theta_k$), integrate over the angle $\theta_k$, rescale the wave vector with the Gouy-Chapman length as $k\to q=\mu k$, and pass to the adimensional separation distance $z_a\to \bz=|z_a|/\mu$.  The self-energy~(\ref{pp4}) takes the form
\be
\label{pp5}
\Delta\Omega_{pp}(\bz)=\frac{\Xi_p}{2}\Phi(\bz),
\ee
where we introduced the electrostatic coupling parameter  $\Xi_p=Q_p^2\ell_B/\mu$ associated with the polymer charge
\be
Q_p=L\tau, 
\ee
and the adimensional self-energy
\bea\label{pp6}
\Phi(\bz)&=&\int_0^{+\infty}\frac{\mathrm{d}q}{q^2}\left\{-\left(1+\bz\right)^{-2}\right.\\
&&\hspace{1.65cm}\left.+\Delta_c(q)\left[\left(1+\bz\right)^{-1}+q\right]^2e^{-2q\bz}\right\}.\nonumber
\eea
In the limit of vanishing dielectric discontinuity $\e_m=\e_w$, the functional form of Eq.~(\ref{pp6}) coincides with the ionic self energy Eq.(44) of Ref.~\cite{Netz2000}. 

\begin{figure}
\includegraphics[width=1.1\linewidth]{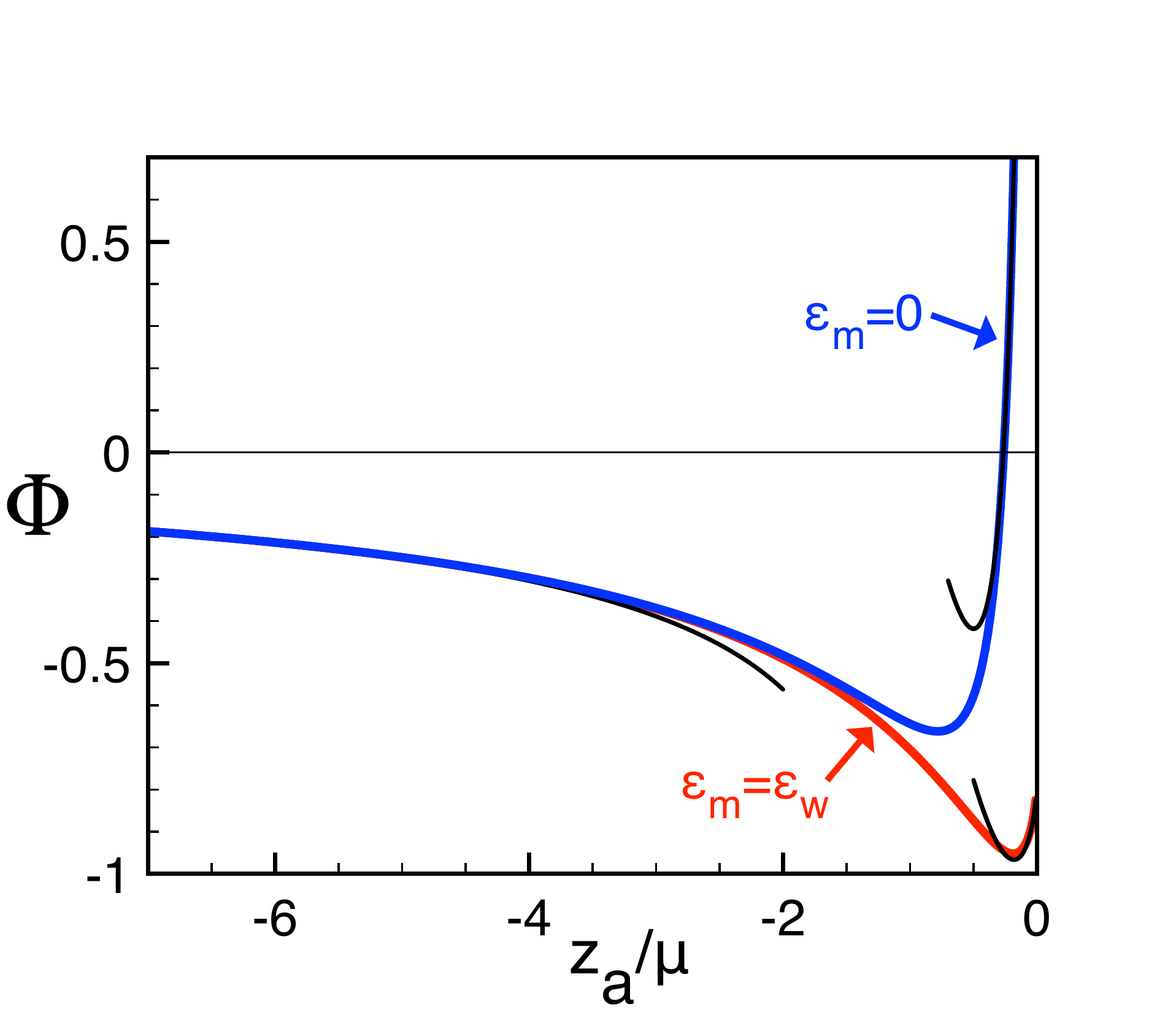}
\caption{(Color online) Adimensional polymer self energy of Eq.~(\ref{pp7}) for $\e_m=\e_w$ (red curve) and Eq.~(\ref{pp10}) for $\e_m=0$ (blue curve). Thin black curves display the limiting laws for very close ($\bz\ll1$) and very large separation distances ($\bz\gg1$) from the membrane surface (see the main text).}
\label{fig2}
\end{figure}

In terms of the same adimensional parameters, the polymer-membrane charge energy~(\ref{pm2}) reads
\be
\label{pm3}
\Omega_{pm}(\bz)=-\frac{2Q_p}{q}\ln(1+\bz).
\ee
Thus in the counterion-only regime, the net polymer grand potential~(\ref{ptot1}) given by
\be\label{ptot2}
\Delta\Omega_p(z_a)=-\frac{2Q_p}{q}\ln(1+\bz)+\frac{\Xi_p}{2}\Phi(\bz)
\ee
depends exclusively on the ratio of the polymer and ion charges $Q_p/q$, the electrostatic coupling parameter $\Xi_p$ that quantifies the deviation from the MF potential~(\ref{pm3}), and the dielectric discontinuity parameter $\eta$. One notes that with increasing distance from the membrane surface, the MF grand potential~(\ref{pm3}) of the polymer drops monotonically. This corresponds to the MF-level similar charge repulsion between the DNA and the membrane charges. In order to consider correlation effects carried by the second term of Eq.~(\ref{ptot2}), we focus first on the limit of vanishing dielectric discontinuity $\e_m=\e_w$. In this limit, the adimensional self-energy~(\ref{pp6}) presents an analytical form. Setting $\eta=1$, one finds 
\bea
\label{pp7}
\Phi(\bz)&=&-\frac{2\bz}{(1+\bz)^2}\\
&&+\frac{e^{(1-i)\bz}}{2(1+\bz)^2}\left\{(\bz-i)^2\left(\pi+i\;\mathrm{Ei}[(-1+i)\bz]\right)\right.\nonumber\\
&&\hspace{1cm}+\left.e^{2i\bz}(\bz+i)^2\left(\pi-i\;\mathrm{Ei}[-(1+i)\bz]\right)\right\},\nonumber
\eea
where $i$ stands for the imaginary unit number and $\mathrm{Ei}(x)$ is the exponential integral function~\cite{math}. In Fig.~\ref{fig2}, we plotted the adimensional polymer self-energy~(\ref{pp7}) (red curve). One notes that the self-energy drops towards the interface and exhibits an attractive minimum in its neighborhood. We emphasize that this peculiarity was also found in the self-energy of counterions~\cite{Netz2000}. The effect is due to the dense surface counterion layer intensifying the screening ability of the interfacial region. Since the screening of the electrostatic field associated with the polymer charge lowers the polymer free energy, this translates into an attractive \textit{solvation} force driving the polyelectrolyte towards the membrane surface. 

At small separation distances from the wall, i.e. for $\bz\ll1$ or $|z_a|\ll\mu$, the adimensional self-energy~(\ref{pp7}) has the asymptotic behavior
\be
\label{pp8}
\Phi(\bz)=-\frac{\pi}{4}+\left\{\frac{1}{2}\ln\left(2\bz^2\right)+\frac{\pi}{4}+\gamma-1\right\}\bz+O\left(\bz^2\right),
\ee
where $\gamma=0.577(2)$ is Euler's constant. The asymptotic law~(\ref{pp8}) reported in Fig.~\ref{fig2} is seen to reproduce accurately the minimum of the self energy. According to Eq.~(\ref{pp8}), the self-energy minimum is located at
\be
\label{minpo}
z_a^*=-\frac{\mu}{\sqrt{2}}\exp{\left(-\frac{\pi}{4}-\gamma\right)}.
\ee
Eq.~(\ref{minpo}) indicates that the larger is the membrane charge, the closer the minimum gets to the membrane surface (i.e. $\sigma_m\uparrow$ $|z_a^*|\downarrow$).  In the opposite limit of large distances $\bz\gg1$, the potential~(\ref{pp7}) dissipates algebraically as
\be
\label{pp9}
\Phi(\bz)=-\frac{3}{2}\left(\frac{1}{\bz}-\frac{1}{\bz^2}+\frac{1}{\bz^3}\right)+O\left(\bz^{-4}\right).
\ee
According to the asymptotic law~(\ref{pp9}), the self-energy is shorter ranged than the MF part~(\ref{pm3}) of the grand potential~(\ref{ptot2}). Hence, far enough from the membrane surface, the polymer immersed in a counterion-only liquid will always experience an overall repulsion.
 
Biological and artificial membranes are usually made of carbon-based materials with low dielectric permittivity $\e_m\sim2$. In order to investigate the resulting image-charge effects, we will consider the close limit of $\e_m=0$ or $\eta\to0$ where the adimensional self-energy~(\ref{pp6}) can be evaluated analytically. In this limit, the self-energy takes the closed-form 
\bea
\label{pp10}
\Phi(\bz)&=&-\frac{(\bz-1)(1+3\bz)}{2\bz(1+\bz)^2}\\
&&+\frac{e^{(1+i\sqrt{3})\bz}}{3(1+\bz)^2}\left\{\pi-i\;\mathrm{Ei}[-(1+i\sqrt{3})\bz]\right\}\nonumber\\
&&\hspace{3mm}\times\left\{-2\sqrt{3}+\bz\left[6i-2\sqrt{3}+(3i+\sqrt{3})\bz\right]\right\}\nonumber\\
&&+\frac{e^{(1-i\sqrt{3})\bz}}{3(1+\bz)^2}\left\{\pi+i\;\mathrm{Ei}[(-1+i\sqrt{3})\bz]\right\}\nonumber\\
&&\hspace{3mm}\times\left\{-2\sqrt{3}-2(3i+\sqrt{3})\bz+(-3i+\sqrt{3})\bz^2\right\}.\nonumber
\eea
Fig.~\ref{fig2} shows that the adimensional self-energy~(\ref{pp10}) (blue curve) embodies two different correlation effects. Namely, the dense counterion layer enhancing the screening ability of the interfacial region results in an attractive potential minimum. This is followed by the high interfacial barrier resulting from polymer-image charge interactions.  Indeed, in the vicinity of the interface $\bz\ll1$, Eq.~(\ref{pp10}) behaves as
\be
\label{pp11}
\Phi(\bz)=\frac{1}{2\bz}-\frac{4\pi}{3\sqrt{3}}+2\bz+O\left(\bz^2\right)
\ee
Eq.~(\ref{pp11}) is reported in Fig.~\ref{fig2} by the thin black curve. One notes that the first term of the asymptotic law~(\ref{pp11}) corresponds to the standard image-charge self-energy. The latter diverges algebraically with the distance from the dielectric membrane surface. In the opposite limit of large separation distances $\bz\gg1$, the potential~(\ref{pp10}) has the same asymptotic law~(\ref{pp9}) as the system without dielectric discontinuity (see also Fig.~\ref{fig2}). Thus, far enough from the interface, attractive solvation forces induced by the counterion cloud always take over the repulsive image-charge interactions. 

We consider next the effect of salt on polymer-membrane interactions.

\subsection{Symmetric electrolytes}

We investigate here correlation effects in symmetric electrolytes.  We focus on the limits of short and long polymers where the technical task is considerably reduced as the double integral of Eq.~(\ref{pp3}) transforms to a simple integral. 

\subsubsection{Short Polymers}
\label{sp}

We consider first the case of short polymers that provides analytically transparent results. As in Section~\ref{countlim}, we Taylor-expand the sinusoidal function of Eq.~(\ref{pp3}) and switch to polar coordinates. The polymer self-energy becomes
\bea
\label{pp14II}
\Delta\Omega_{pp}(z_a)&=&\frac{Q_p^2\ell_B\kappa_b^2}{2}\int_0^{\infty}\frac{\mathrm{d}k}{pk}\left\{-\mathrm{csch}^2\left[\kappa_b(z_a-z_0)\right]\right.\nonumber\\
&&\hspace{0cm}\left.+\Delta(k)\left(\frac{p}{\kappa_b}-\mathrm{coth}\left[\kappa_b(z_a-z_0)\right]\right)^2e^{2pz_a}\right\}.\nonumber\\
\eea
We note that the integral of Eq.~(\ref{pp14II}) has the functional form of the ionic self energy calculated in Ref.~\cite{Buyuk2012}. Changing now the integration variable as $k\to u=p/\kappa_b$,  and introducing the rescaled separation distance $\tz=-\kappa_bz_a$ and the coupling parameter 
\be
\Gamma_p=Q_p^2\kappa_b\ell_B,
\ee
the polymer self-energy takes the form
\bea
\label{pp15}
\Delta\Omega_{pp}(\tz)&=&\frac{\Gamma_p}{2}\Theta(\tz),
\eea
with the dimensionless self-energy
\bea
\label{pp16}
\Theta(\tz)&=&\int_{1}^{+\infty}\frac{\mathrm{d}u}{u^2-1}\left\{-\mathrm{csch}^2\left[\ln(\gamma_c)-\tz\right]\right.\\
&&\hspace{0cm}\left.+\tdel(u)\left(u-\mathrm{coth}\left[\ln(\gamma_c)-\tz\right]\right)^2e^{-2u\tz}\right\}.\nonumber
\eea
In Eq.~(\ref{pp16}), we introduced the function
\be\label{del2}
\tdel(u)=\frac{1+\left(u-\eta\sqrt{u^2-1}\right)s\left(su-\sqrt{s^2+1}\right)}{1+\left(u+\eta\sqrt{u^2-1}\right)s\left(su+\sqrt{s^2+1}\right)}.
\ee
In terms of the same adimensional parameters, the MF-level polymer-membrane coupling energy~(\ref{pm1}) reads
\be
\label{pm4}
\Omega_{pm}(\tz)=\frac{2Q_p}{q}\ln\left[\frac{1+\gamma_c\;e^{-\tz}}{1-\gamma_c\;e^{-\tz}}\right].
\ee
Hence, the polymer grand potential~(\ref{np}) is given by
\be\label{pp16II}
\Delta\Omega_p(\tz)=\frac{2Q_p}{q}\ln\left[\frac{1+\gamma_c\;e^{-\tz}}{1-\gamma_c\;e^{-\tz}}\right]+\frac{\Gamma_p}{2}\Theta(\tz).
\ee
Comparing the first and the second terms of Eq.~(\ref{pp16II}), one notes that the relative weight of charge correlations scales as $Q_p\kappa_b$. Thus, for short polymers, charge correlations are amplified with increasing salt concentration or polymer charge.

\begin{figure}
\includegraphics[width=1.\linewidth]{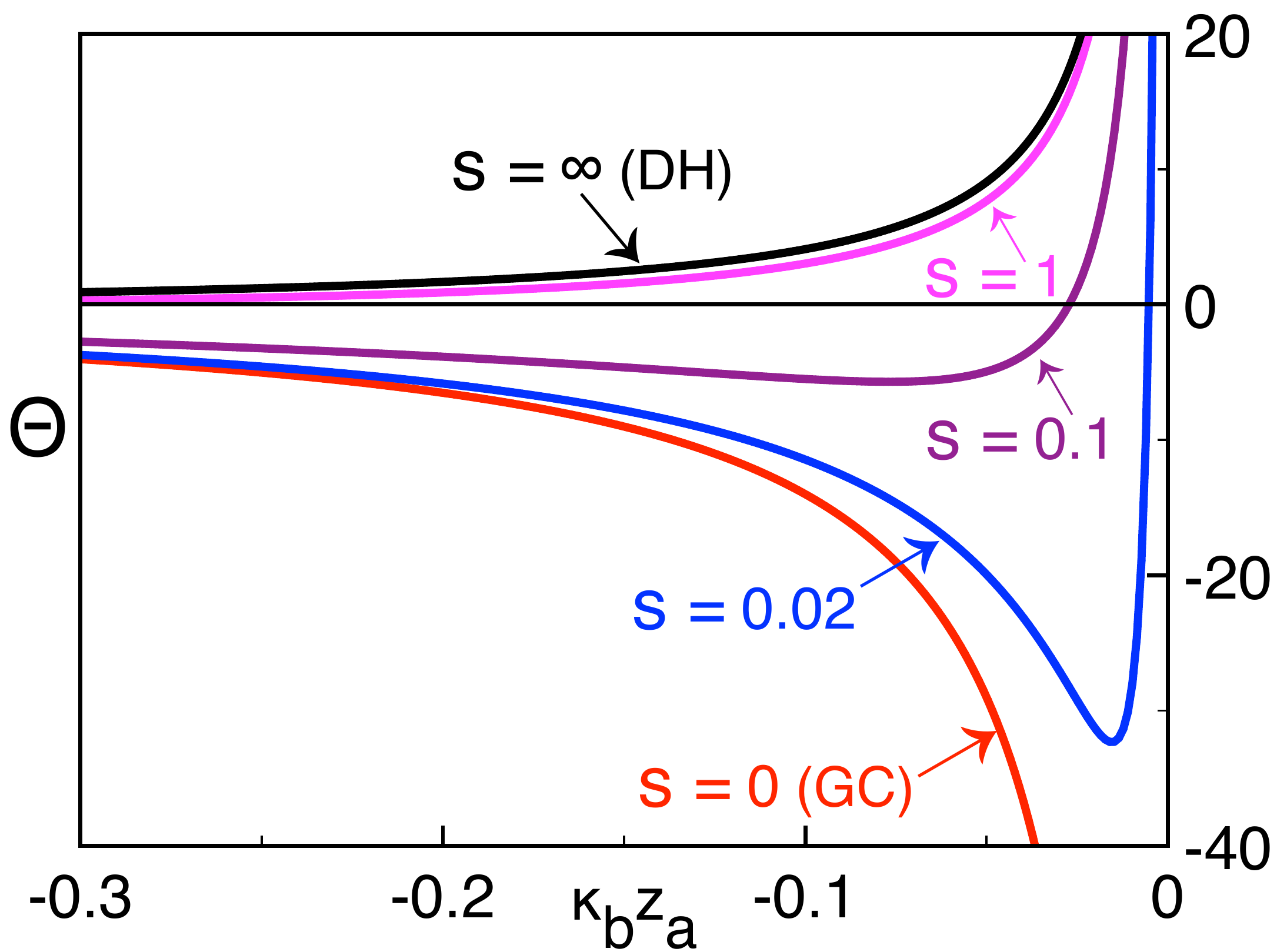}
\caption{(Color online) Dimensionless self-energy of short polymers~(\ref{pp16}) versus the distance from the membrane with permittivity $\e_m=2$ at different values of the parameter $s=\kappa_b\mu$.} 
\label{fig3}
\end{figure}

In Fig.~\ref{fig3}, we plotted the self-energy~(\ref{pp16}) by changing the adimensional parameter $s=\kappa_b\mu$. We note that the lower is the parameter $s$,  the lower is the salt density ($\rho_b\downarrow s\downarrow$) or the stronger is the membrane charge ($\sigma_m\uparrow s\downarrow $).  Decreasing the parameter $s$ (i.e. rising the membrane charge or lowering the salt concentration), the screening ability of the interfacial region is amplified. As a result, Fig.~\ref{fig3} shows that the self-energy switches from repulsive to attractive. Thus, unlike counterion-only liquids where charge correlations always bring an attractive contribution (see Fig.~\ref{fig2}), in the presence of salt, the role played by correlations can be totally reversed by tuning the salt density.

In order to evaluate the correction brought by a finite membrane charge to the DH limit of neutral membranes $\sigma_m=0$ (or $s=\infty$), we evaluate the Taylor expansion of the self-energy~(\ref{pp16}) at the order $O\left(s^{-2}\right)$.  To account for the dielectric jump between the membrane and the electrolyte, we fixe the membrane permittivity to $\e_m=0$ (i.e. $\eta=0$). Carrying out the Fourier integral, one finds that the self-energy decomposes as $\Theta(\tz)\approx\Theta^{(0)}_{DH}(\tz)+s^{-2}\Theta^{(1)}_{DH}(\tz)$. The DH part given by
\be
\label{pp17}
\Theta^{(0)}_{DH}(\tz)=\frac{e^{-2\tz}}{2\tz}
\ee
corresponds to screened image-polymer charge interactions repelling the polymer from the membrane (the black curve in Fig.~\ref{fig3})~\cite{Netz1999,Buyuk2012}. The correction term reads in turn
\bea\label{pp18}
\Theta^{(1)}_{DH}(\tz)&=&-\frac{1}{2}\left\{\left[\gamma+\ln(4\tz)\right]e^{-2\tz}-4\;\mathrm{Ei}(-2\tz)\right.\nonumber\\
&&\left.\hspace{7mm}+\left(2+e^{2\tz}\right)\mathrm{Ei}(-4\tz)\right\}.
\eea
In the neighbourhood of the interface or at weak salt $\tz\ll1$, the correction term rises linearly with distance as
\be
\label{pp19}
\Theta^{(1)}_{DH}(\tz)\approx-\ln(4)+2\tz.
\ee
At large distances from the interface or in strong salt solutions $\tz\gg1$, Eq.~(\ref{pp18})  exhibits an exponential decay,
\be
\label{pp20}
\Theta^{(1)}_{DH}(\tz)\approx-\frac{1}{2}e^{-2\tz}\left\{\gamma+\ln(4\tz)+\frac{7}{4\tz}\right\}.
\ee
First of all, in both regimes, the finite-charge correction~(\ref{pp18})  associated with the enhanced interfacial charge screening is negative. This explains the reduction of the DH potential by a finite membrane charge density in Fig.~\ref{fig3}. Secondly, one notes that at large separation distances $\tz\gg1$, the correction term~(\ref{pp20}) dominates the DH potential~(\ref{pp17}). Thus, far enough from the surface of the charged membrane, charge correlations  always make an attractive contribution to polymer-membrane interactions.

We focus now on the opposite Gouy-Chapman (GC) regime $s\ll1$ of strongly charged membranes or dilute electrolytes. By Taylor-expanding the self-energy~(\ref{pp16}) at the order $O\left(s\right)$, one gets $\Theta(\tz)\approx\Theta^{(0)}_{GC}(\tz)+s\Theta^{(1)}_{GC}(\tz)$. The self-energy of the strict GC limit $s\to0$ reads
\bea
\label{pp21}
\Theta^{(0)}_{GC}(\tz)&=&\frac{\mathrm{csch}^2(\tz)}{8\tz}\left\{4\tz\left[-\gamma+\mathrm{Ei}(-4\tz)-\ln(4\tz)\right]\right.\nonumber\\
&&\hspace{1.5cm}\left.+\left(1-e^{-2\tz}\right)^2\right\},
\eea
while the correction term is given by
\bea
\label{pp22}
\Theta^{(1)}_{GC}(\tz)&=&\frac{e^{-2\tz}}{2\tz^2}\left\{-1-2\tz-4\tz\coth(\tz)\right.\\
&&\hspace{1.cm}+2\tz^2\coth(\tz)\left[1+\coth(\tz)\right]^2\nonumber\\
&&\hspace{1.2cm}\left.\times\left[\gamma+\ln(4\tz)-\mathrm{Ei}(-4\tz)\right]\right\}.\nonumber
\eea
The GC potential~(\ref{pp21}) reported in Fig.~\ref{fig3} by the red curve is seen to drop without lower bound. Indeed, in the vicinity of the interface $\tz\ll1$, this potential that accounts for the enhanced screening ability of the interfacial region diverges as
\be
\label{pp23}
\Theta^{(0)}_{GC}(\tz)\approx-\frac{3}{2\tz}+1-\frac{\tz}{9}.
\ee
Thus, the GC limit of the self-energy is purely attractive. The image-charge barrier associated with the dielectric discontinuity is in turn included in the correction term~(\ref{pp22}) that exhibits the asymptotic divergence 
\be
\label{pp24}
\Theta^{(1)}_{GC}(\tz)\approx\frac{3}{2\tz^2}-\frac{1}{9}
\ee
at $\tz\ll1$. In Fig.~\ref{fig3},  the asymptotic behaviors~(\ref{pp23}) and~(\ref{pp24}) correspond respectively to the decreasing and increasing branches of the self-energy curves. From the close distance limit of the GC-expansion $\Theta(\tz)\approx\Theta^{(0)}_{GC}(\tz)+s\Theta^{(1)}_{GC}(\tz)$, one finds that the minimum of the self energy curves that joins these two branches is located at $|z_a^*|\approx\mu$.  Thus, in agreement with Eq.~(\ref{minpo}) for the counterion-only liquid, the larger is the membrane charge, the closer the potential minimum gets to the wall.

\begin{figure}
\includegraphics[width=1.1\linewidth]{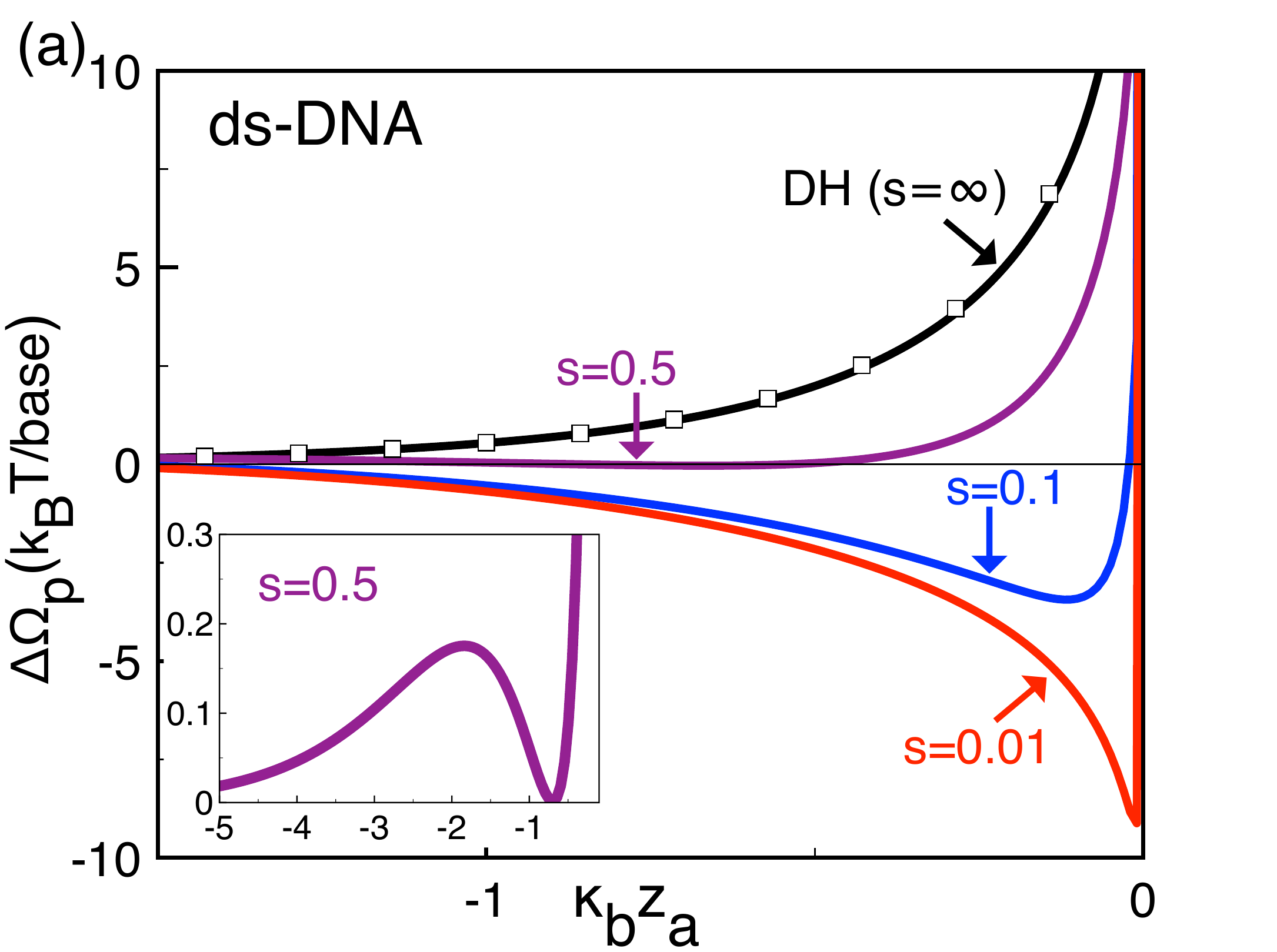}
\includegraphics[width=1.1\linewidth]{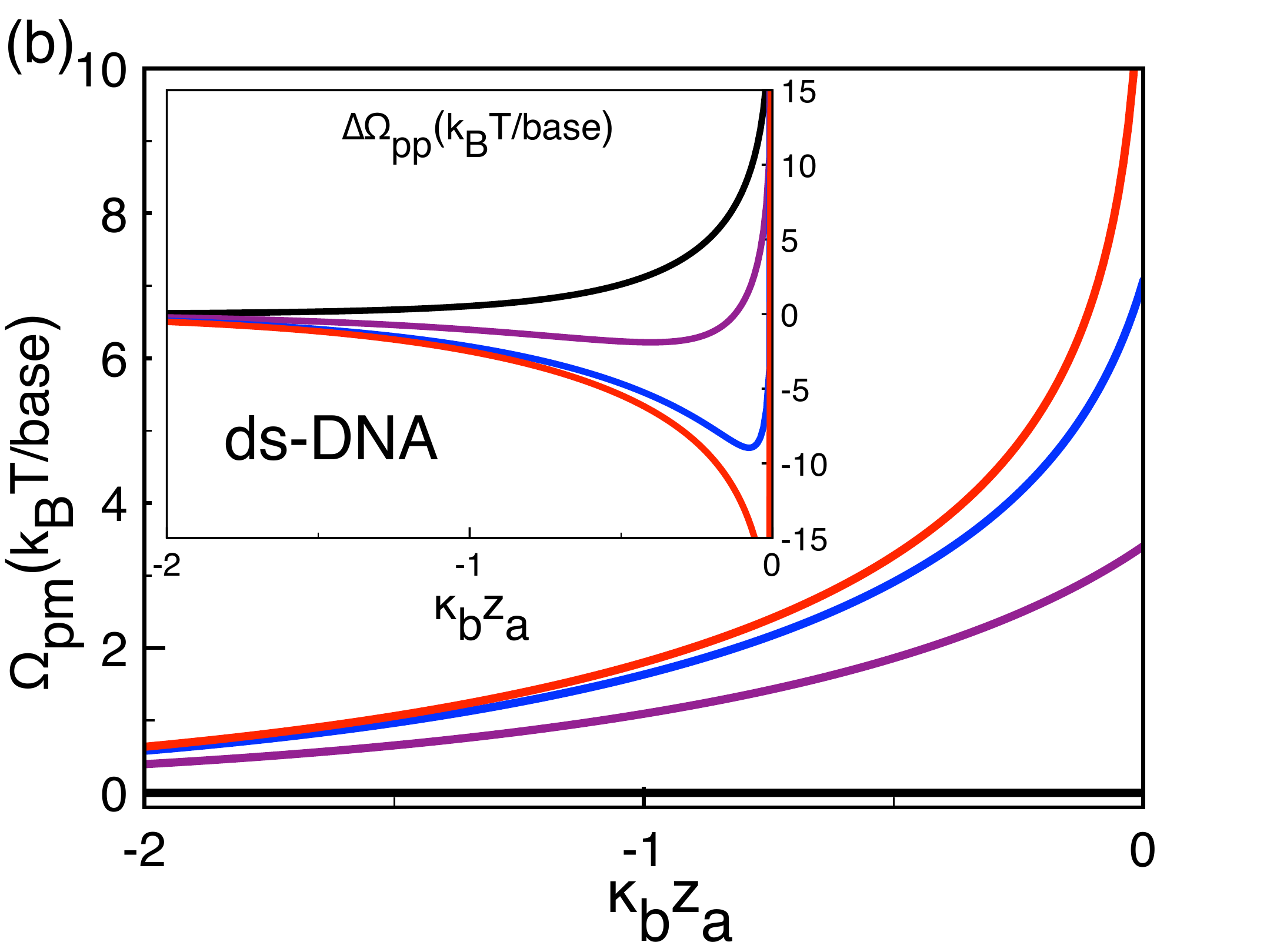}
\includegraphics[width=1.1\linewidth]{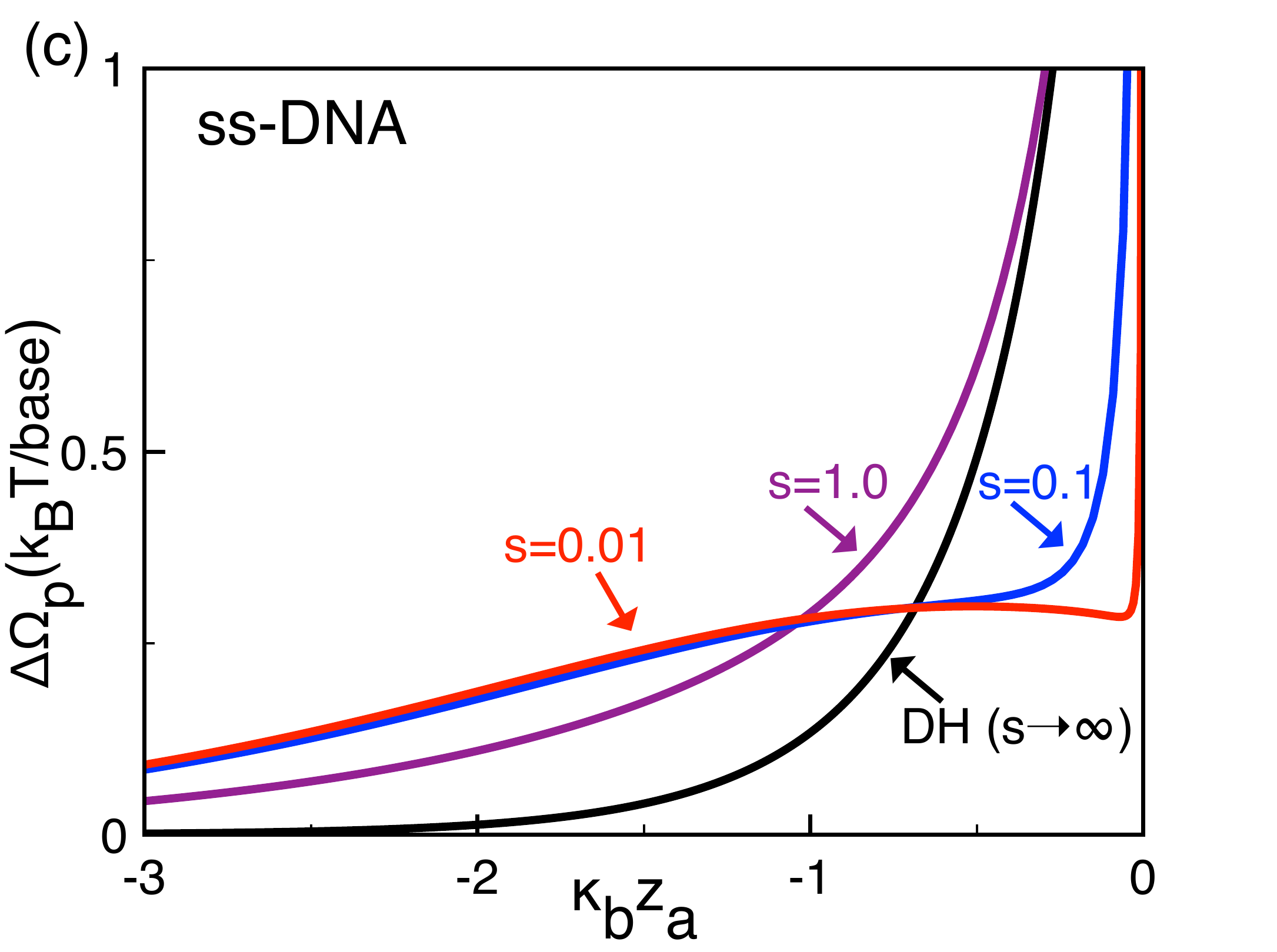}
\caption{(Color online) (a) One-loop polymer grand potential~(\ref{pp14}),  (b) MF-level grand potential~(\ref{pm4}) (main plot) and self-energy~(\ref{pp12}) (inset) of a ds-DNA with linear charge density  $\tau=0.59$ $e/${\AA}. (c) displays the total grand potential of a ss-DNA molecule with charge density  $\tau=0.29$ $e/${\AA}. The membrane permittivity is $\e_m=2$ (solid curves). The square symbols in (a) correspond to the DH regime at the membrane permittivity $\e_m=0$ where the self-energy~(\ref{pp13}) diverges logarithmically as $\Psi(\tz)=K_0(2\tz)\simeq-\ln(\tz)$. The inset in (a) zooms to the coexistence region.} 
\label{fig4}
\end{figure}

\subsubsection{Long polymers and similar charge attraction}
\label{lp}

In light of the analysis of charge correlations considered in the previous part, we investigate now the biologically relevant regime of long polymers. Taking the limit $L\to\infty$,  the sinusoidal function in Eq.~(\ref{pp3}) yields a Dirac delta distribution (see Eq.~(\ref{dir})), which cancels the integral on the wavevector $k_x$. Passing to the adimensional integration variable $k_y\to u=\sqrt{1+(k_y/\kappa_b)^2}$, the grand potential~(\ref{pp3}) becomes
\bea
\label{pp12}
\Delta\Omega_{pp}(\tz)&=&L\ell_B\tau^2\Psi(\tz),
\eea
with the dimensionless self-energy
\bea
\label{pp13}
\Psi(\tz)&=&\int_{1}^{+\infty}\frac{\mathrm{d}u}{(u^2-1)^{3/2}}\left\{-\mathrm{csch}^2\left[\ln(\gamma_c)-\tz\right]\right.\\
&&\hspace{1.1cm}\left.+\tdel(u)\left(u-\mathrm{coth}\left[\ln(\gamma_c)-\tz\right]\right)^2e^{-2u\tz}\right\}.\nonumber
\eea
Thus, the total polymer grand potential~(\ref{np}) is given by
\be\label{pp14}
\Delta\Omega_p(\tz)=\frac{2Q_p}{q}\ln\left[\frac{1+\gamma_c\;e^{-\tz}}{1-\gamma_c\;e^{-\tz}}\right]+L\ell_B\tau^2\Psi(\tz).
\ee
First, one notes that for long polymers, the total grand potential~(\ref{pp14}) scales linearly with the polymer length $L$. Then,  the dependence of the potential~(\ref{pp14}) on the membrane charge $\sigma_m$ and salt density $\rho_b$ is solely encoded in the parameter $s=\kappa_b\mu$. Thus, for long polymers, one can explore the whole surface charge and ion concentration regime by changing exclusively the parameter $s$. 

Fig.~\ref{fig4}(a) displays the total grand potential~(\ref{pp14}), and Fig.~\ref{fig4}(b) shows the MF-level polymer-membrane charge coupling energy~(\ref{pm4}) (main plot) and the polymer self-energy~(\ref{pp12}) (inset) at different values the parameter $s$.  The membrane permittivity is $\e_m=2$. The polymer charge density is fixed to the value of \textit{ds-DNA} molecules $\tau=0.59$ $e/${\AA}. In the DH-limit $s\to\infty$ of neutral membranes (black curves) and close to the membrane surface, the polymer encounters a repulsive barrier (Fig.~\ref{fig4}(a)). In Fig.~\ref{fig4}(b), one notes that this effect is solely due to the polymer self-energy embodying the repulsive polymer-image charge interaction. Indeed, at the membrane permittivity $\e_m=0$ (square symbols),  the adimensional self-energy~(\ref{pp13}) has the analytical form $\Psi(\tz)=K_0(2\tz)$ exhibiting a logarithmic divergence $\Psi(\tz)\simeq-\ln(\tz)$ at the interface~\cite{Netz2000}.

Decreasing the parameter $s$ (i.e. reducing the salt concentration or rising the membrane charge),  the MF-level membrane-DNA repulsion energy $\Omega_{pm}$ is enhanced (the main plot of Fig.~\ref{fig4}(b)). Furthermore, below the value $s\simeq1$ where one gets into the GC regime characterized in the previous section, the DNA self-energy $\Delta\Omega_{pp}$ turns from repulsive to attractive (inset).  At the \textit{ds-DNA} charge considered in Figs.~\ref{fig4}(a) and (b), the weight of the self-energy~(\ref{pp12}) dominates the MF grand potential~(\ref{pm4}). Consequently, the total grand potential develops an attractive well that becomes deeper with decreasing $s$. In other words, a larger negative membrane charge results in a stronger attraction of the negatively charged ds-DNA molecule. This correlation-induced like-charge attraction effect is one of the key results of our work. In Fig.~\ref{fig4}(c), we consider now  the grand potential~(\ref{pp14})  of \textit{ss-DNA} molecules having a weaker charge density $\tau=0.29$ $e/${\AA}. One notes that although the interfacial barrier is lowered with decreasing $s$, the grand potential does not develop a stable attractive well. We verified that below the value $s=0.01$ (red curve), the grand potential profile is practically unchanged.  Thus, for ss-DNA molecules, the enhanced interfacial screening of the DNA charges is not strong enough to turn the DNA-membrane interaction from repulsive to attractive.

\begin{figure}
\includegraphics[width=1.0\linewidth]{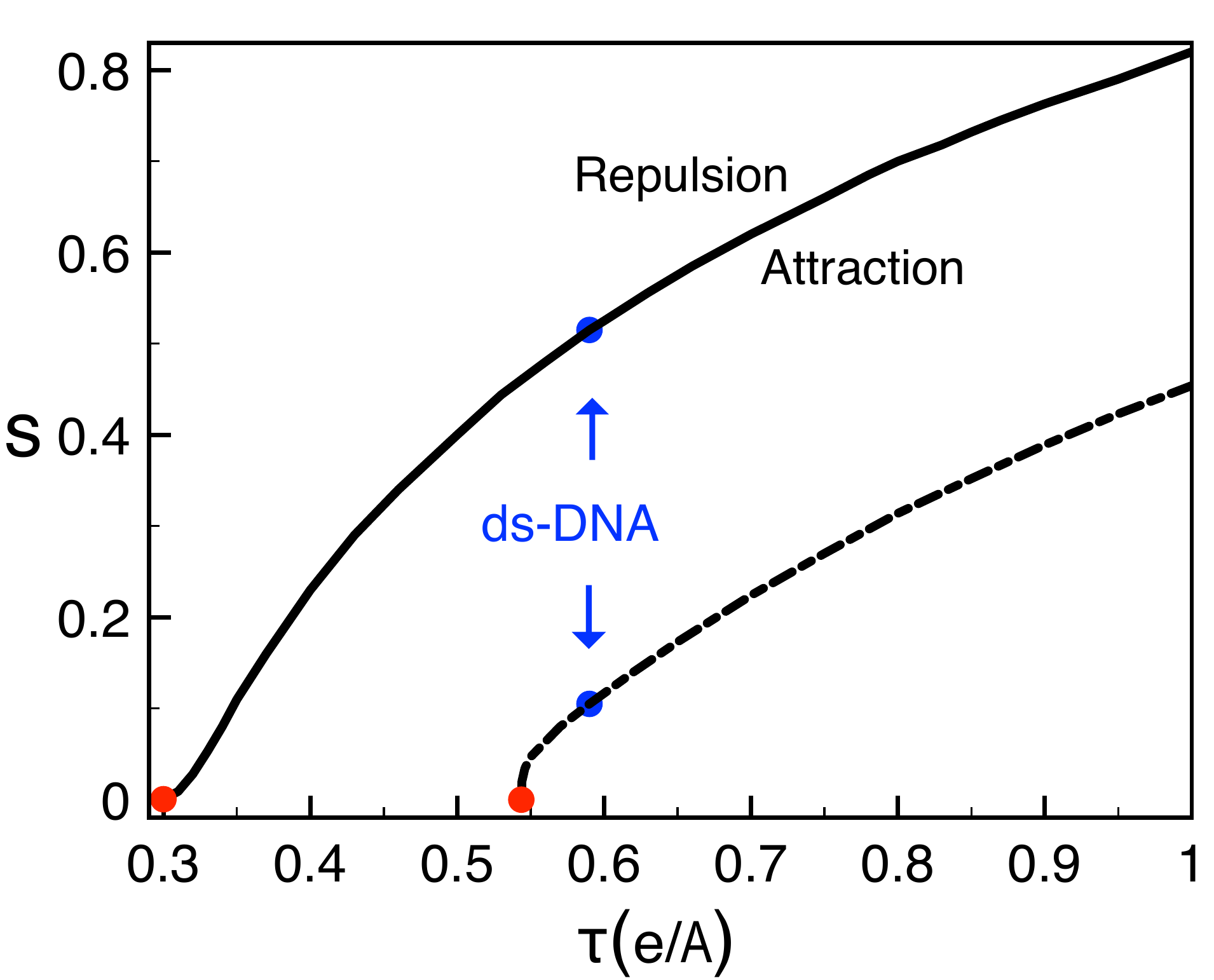}
\caption{(Color online) Phase diagram : Critical values of the parameter $s=\kappa_b\mu$ versus the polymer charge density $\tau$ separating the parameter regimes with binding (area below the critical lines) and unbinding polyelectrolytes (area above the lines) oriented parallel (solid curve) and perpendicular to the membrane surface (dashed curve). The red dots mark the characteristic polymer charge densities reached at $s=0$ below which the adsorption state disappears.} 
\label{fig5}
\end{figure}

In Fig.~\ref{fig5}, we plotted the phase diagram (solid curve) that displays the critical values of the parameter $s=s^*=\kappa_b^*\mu^*$ versus the polymer charge density $\tau$ separating the parameter regimes with attractive and repulsive membranes. The critical line corresponds to the phase coexistence where the grand potential exhibits a minimum at the interface (adsorption state) and a second minimum in the bulk (desorption state). In the case of ds-DNA molecules, this corresponds to the value $s=0.5$ (see the inset of Fig.~\ref{fig4}(a) and the blue dot in Fig.~\ref{fig5}). For polyethylene terephthalate membranes with surface charge $\sigma_m=0.16$ $C/m^2$ at $pH=7$, this value can be reached at the salt concentration $\rho_b=0.5$ M. First, the diagram of Fig.~\ref{fig5} shows that the weaker is the polyelectrolyte charge, the higher should be the membrane charge (or the lower the salt density) for the polymer adsorption to occur (i.e. $\tau\downarrow\sigma_m^*\uparrow$ or $\tau\downarrow\rho_b^*\downarrow$). This can be explained by the fact that a lower polymer charge has to be compensated by a stronger counterion excess for the enhanced interfacial screening to result in the attraction of the polymer. Then, one notes that the critical line ends at the characteristic polymer charge density $\tau_c=0.3$ $e/${\AA}  reached in the GC limit $s=0$ (red dot) below which the adsorption state disappears. In agreement with Fig.~\ref{fig4}(c), this means that ss-DNA molecules should be always repelled by the negatively charged membrane, regardless of the membrane charge and the salt density. Finally, the diagram of Fig.~\ref{fig5} indicates that the increase of the bulk salt concentration should result in the unbinding of a polyelectrolyte initially adsorbed to the similarly charged membrane. The predictions of this phase diagram can be verified by current DNA transport experiments. 

We consider next the effect of correlations on the interaction between a membrane and a polyelectrolyte oriented perpendicular to the membrane wall.

\section{Approach of the polymer prior to adsorption}
\label{app}

We calculate now the grand potential of the polymer oriented perpendicular to the membrane surface. This configuration corresponds to the approach phase of electrophoretically driven DNA molecules prior to translocation events~\cite{e1,e2,e3,e4,e5,e6}. We choose the Cartesian coordinates of the approaching polymer as $(0,0,z_t)$, where the variable $z_t<0$ denotes the distance of the right end of the polymer from the wall (see Fig.~\ref{fig1}). In this configuration, the charge density function is
\be
\label{polchA}
\sigma_p(\br)=-\tau\;\delta(\br_\pa)\;\theta(-z)\;\theta(z_t-z)\;\theta(z-z_t+L).
\ee
Injecting the structure factor~(\ref{polchA}) with the membrane potential~(\ref{EqIV}) into Eq.~(\ref{Eq0II}), the polymer-membrane coupling potential follows as
\bea
\label{pm5}
\Omega_{pm}(z_t)&=&\frac{2\tau}{q\kappa_b}\left\{\mathrm{Li}_2\left[e^{\kappa_b(z_t-z_0)}\right]-\mathrm{Li}_2\left[-e^{\kappa_b(z_t-z_0)}\right]\right.\nonumber\\
&&\left.-\mathrm{Li}_2\left[e^{\kappa_b(z_t-L-z_0)}\right]+\mathrm{Li}_2\left[-e^{\kappa_b(z_t-L-z_0)}\right]\right\},\nonumber\\
\eea
where we used the polylogarithm function $\mathrm{Li}_2(x)$~\cite{math}.

Because of the polymer correlations perpendicular to the dielectric wall, the calculation of the DNA self-energy is involved. Substituting the density function~(\ref{polchA}) together with the Green's function~(\ref{tv1}) into Eq.~(\ref{Eq0I}), the self-energy takes the form
\bea
\label{s1}
\Omega_{pp}(z_t)&=&\frac{\ell_B\tau^2}{2}\int_0^{\infty}\frac{\mathrm{d}kp}{k}\int_{z_t-L}^{z_t}\mathrm{d}z\\
&&\times\left\{\left[h_+(z)+\Delta(k)h_-(z)\right]\int_{z_t-L}^z\mathrm{d}z'h_-(z')\right.\nonumber\\
&&\hspace{5mm}\left.+h_-(z)\int_z^{z_t}\mathrm{d}z'\left[h_+(z')+\Delta(k)h_-(z')\right]\right\}.\nonumber
\eea
In the following parts, we will investigate the energies~(\ref{pm5}) and~(\ref{s1}) for a counterion-only liquid and a symmetric electrolyte. 

\subsection{Counterion liquids}
\label{coon}

We take now the counterion-only limit $\kappa_b\to0$ where the MF grand potential~(\ref{pm5}) becomes
\bea
\label{pm6}
\Omega_{pm}(v)&=&\frac{2Q_p}{q}\left[(v+\bL^{-1})\ln(v+\bL^{-1})\right.\\
&&\left.\hspace{1cm}-(v+\bL^{-1}+1)\ln(v+\bL^{-1}+1)\right].\nonumber
\eea
In Eq.~(\ref{pm6}), we introduced the dimensionless distance $v=-z_t/L$ and polymer length $\bL=L/\mu$. In the same limit $\kappa_b\to0$, the homogeneous solutions~(\ref{solpar}) to the 1l-level kernel equation~(\ref{forker}) reduce to
\be\label{homco}
h_{\pm}(z)=e^{\mp kz}\left[1\pm\frac{1}{k(z-\mu)}\right].
\ee
Inserting Eq.~(\ref{homco}) into the self-energy function~(\ref{s1}), subtracting the bulk grand potential, and passing to the dimensionless integration variable $t=L k$, after long but straightforward algebra, the net self-energy takes the single integral form
\be\label{pp25}
\Delta\Omega_{pp}(v)=\frac{L\ell_B\tau^2}{2}\chi(v),
\ee
with the dimensionless polymer self-energy
\be\label{pp26}
\chi(v)=\int_0^\infty\frac{\mathrm{d}t}{t^2}\left[2F_1(t)+\Delta_c(t/\bL)F_2(t)\right].
\ee
In Eq.~(\ref{pp26}), we used the dielectric jump function~(\ref{delc}) and introduced the auxiliary functions
\begin{widetext}
\bea\label{f1}
F_1(t)&=&e^{t(v+\bL^{-1})}\mathrm{Ei}\left[-t(v+\bL^{-1})\right]-e^{t(v+\bL^{-1}+1)}\mathrm{Ei}\left[-t(v+\bL^{-1}+1)\right]+1-e^{-t}\\
&&-\left\{e^{-t(v+\bL^{-1}+1)}-\mathrm{Ei}\left[-t(v+\bL^{-1}+1)\right]\right\}\left\{-e^{t(v+\bL^{-1})}+\mathrm{Ei}\left[t(v+\bL^{-1})\right]+e^{t(\bL^{-1}+v+1)}-\mathrm{Ei}\left[t(v+\bL^{-1}+1)\right]\right\}\nonumber\\
&&-\MeijerG[\Bigg]{3}{1}{2}{3}{0,1}{0,0,0}{t(v+\bL^{-1})}+\MeijerG[\Bigg]{3}{1}{2}{3}{0,1}{0,0,0}{t(v+\bL^{-1}+1)}\nonumber\\
\label{f2}
F_2(t)&=&\left\{e^{-tv}-e^{t/\bL}\;\mathrm{Ei}\left[-t(v+\bL^{-1})\right]-e^{-t(v+1)}+e^{t/\bL}\;\mathrm{Ei}\left[-t(v+\bL^{-1}+1)\right]\right\}^2,
\eea
\end{widetext}
with the Meijer-G functions $\MeijerG*{m}{n}{p}{q}{a_1, \dots, a_p}{b_1, \dots, b_q}{x}$~\cite{math}.  

\begin{figure}
\includegraphics[width=1.2\linewidth]{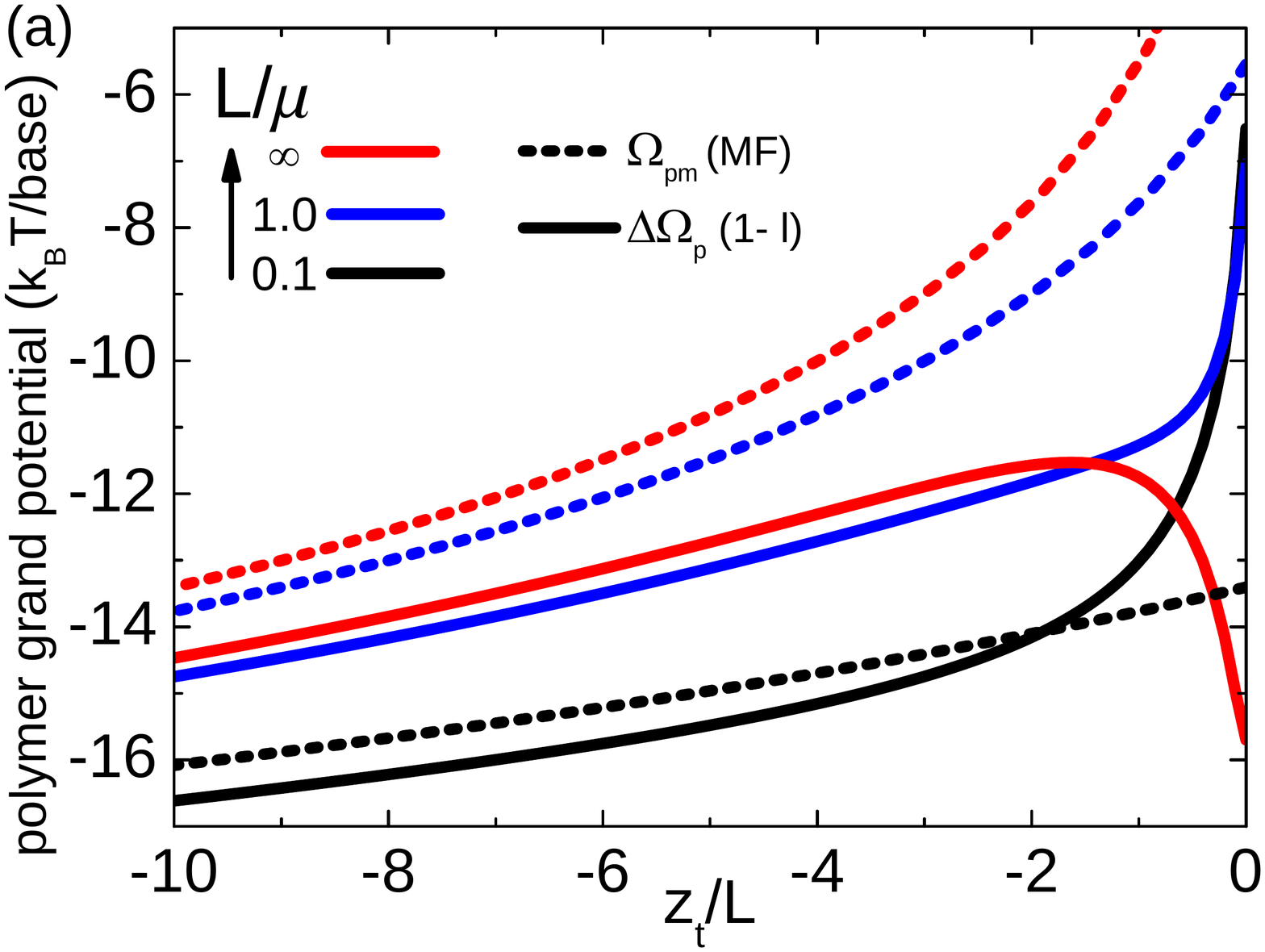}
\includegraphics[width=1.05\linewidth]{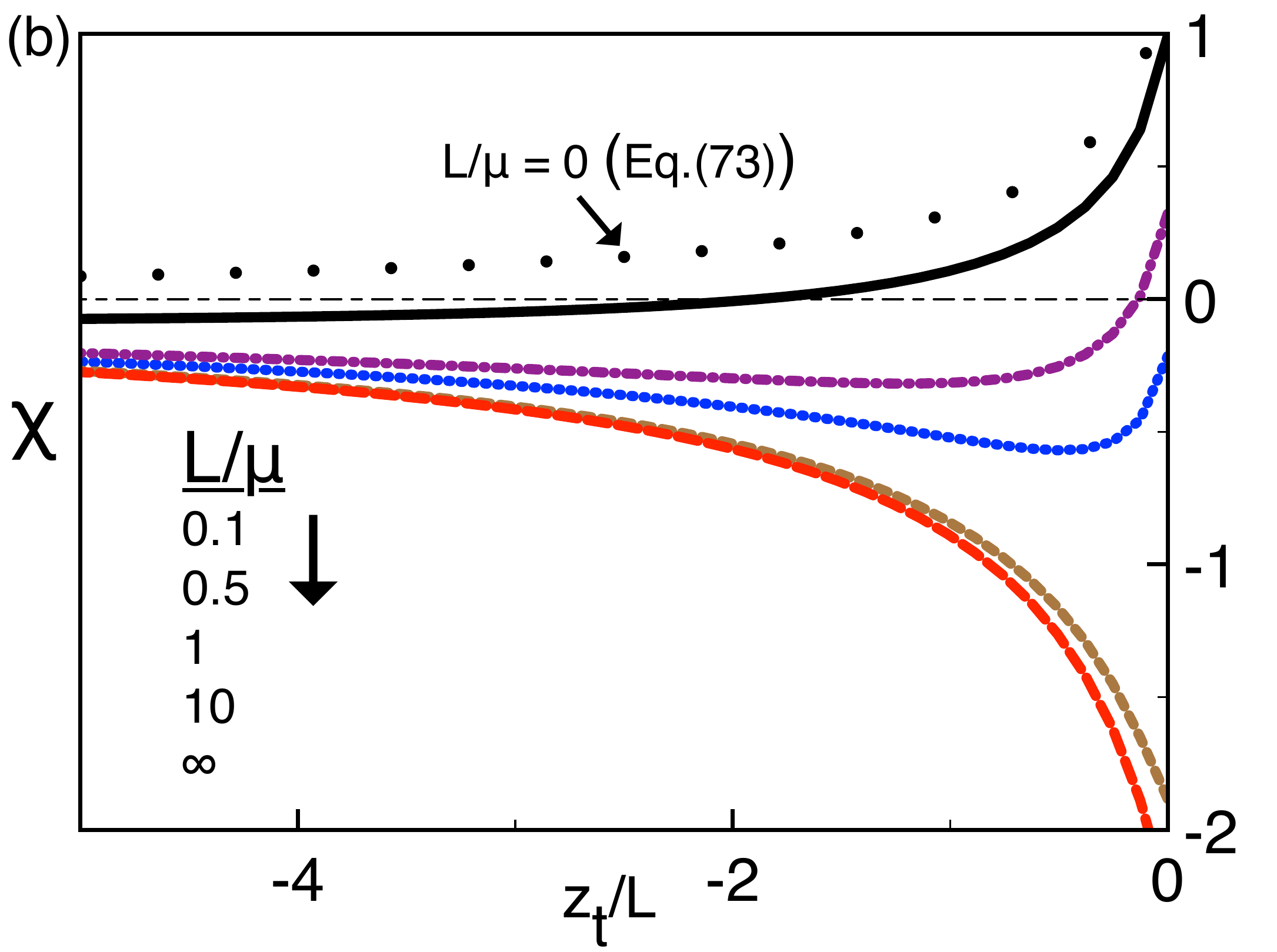}
\caption{(Color online) (a) MF grand potential $\Omega_{pm}(v)$ of Eq.~(\ref{pm6}) (dashed curves), one-loop level grand potential $\Delta\Omega_{p}(v)=\Omega_{pm}(v)+L\ell_B\tau^2\chi(v)/2$ (solid curves), and (b) dimensionless self-energy $\chi(v)$ of Eq.~(\ref{pp26}) versus the reduced separation distance. The membrane permittivity is $\e_m=2$, the polymer charge density $\tau=1.0$ $e/${\AA},  and the different values of the rescaled length $\bL=L/\mu$ are indicated in the legend.} 
\label{fig6}
\end{figure}

We emphasize that at fixed polymer charge density $\tau$, the MF grand potential~(\ref{pm6}) and the self-energy~(\ref{pp25}) per length are solely characterized by the adimensional polymer length $\bL=L/\mu$.  This also implies that the variation of the polymer length $L$ and the membrane charge $\sigma_m$ have the same effect on the polymer grand potential.  In Fig.~\ref{fig6}(a), we show that a larger membrane charge or polymer length results in a more repulsive MF grand potential $\Omega_{pm}(z_t)$ (dashed curves from bottom to top). As a result of the enhanced MF-level repulsion, outside the interfacial region $v\gtrsim1$, the one-loop grand potential $\Delta\Omega_{p}(z_t)=\Omega_{pm}(z_t)+\Delta\Omega_{pp}(z_t)$ rises with the parameter $\bL$ (solid curves). However, this effect is reversed in the interfacial region $v\lesssim1$ where the increase of $\bL$ is accompanied with the drop of the one-loop grand potential. Most importantly, at polymer lengths largely exceeding the GC length $L\gg\mu$, correlations result in an attractive metastable minimum of the total grand potential (the solid red curve for $\bL\to\infty$). Indeed, in counterion-only liquids where the MF grand potential~(\ref{pm6}) and consequently the total grand potential $\Delta\Omega_p$ drops with the distance $z_t$ without lower bound, the repulsion always corresponds to the stable state of the system. We also verified that at the ss-DNA and ds-DNA charge densities, the metastable minimum does not appear.  

In order to explain these points, in Fig.~\ref{fig6}(b), we plotted  the adimensional self-energy~(\ref{pp26}) accounting for charge correlations.  At weakly charged membranes or for short polymers (black curve for $\bL=0.1$), image-charge effects result in a purely repulsive self-energy, explaining the interfacial increase of the MF grand potential by correlations in Fig.~\ref{fig6}(a). Indeed, in the strict limit of neutral membranes $\bL\to0$, the dimensionless polymer self energy~(\ref{pp26}) reduces to the simple form derived in Ref.~\cite{Buyuk2016},
\be
\label{pp28}
\chi(v)=2\Delta_0\left\{\ln\left[\frac{2+2v}{1+2v}\right]+v\ln\left[\frac{4v(1+v)}{(1+2v)^2}\right]\right\},
\ee
with the unscreened dielectric jump function $\Delta_0=(1-\eta)/(1+\eta)$. The self energy reported in Fig.~\ref{fig6}(b) by black dots is seen to be purely repulsive. At the interface, it tends to the finite value $\chi(0)=2\ln(2)$. At separation distances much larger than the polymer length $v\gg1$ (or $|z_t|\gg L$), the potential~(\ref{pp28}) reduces to the unscreened ionic image-charge potential $\chi(v)=\Delta_0/(2v)$. Increasing now the polymer length or the membrane charge from top to bottom, an attractive potential minimum sets in at $L\sim\mu$. At larger polymer lengths $L\gg\mu$, the self-energy drops and becomes purely attractive. This peculiarity due again to the enhanced interfacial charge screening is responsible for the reduction of the MF grand potential and the presence of a metastable minimum in Fig.~\ref{fig6}(a). 

In the next subsection we deal with the effect of salt on the approach phase of translocating DNA molecules.

\subsection{Symmetric electrolytes}

In this part, we calculate the self-energy of the polymer in a symmetric electrolyte and oriented perpendicular to the membrane surface. Because of the non-linear dependence of the homogeneous solutions~(\ref{solpar}) on the coordinate $z$, the spatial integrals \textcolor{black}{of} the self energy Eq.~(\ref{s1}) cannot be evaluated analytically. In order to overcome this complication, we Taylor-expand the functions~(\ref{solpar}) in terms of the parameter $\gamma_c(s)=e^{-\kappa_b z_0}$ as
\be
\label{homt}
h_\mp(z)=\frac{\kappa_b}{p}\sum_{n\geq0}b_n^{\pm}e^{v_n^\pm \kappa_bz},
\ee
where we passed to the adimensional wavevector $u=p/\kappa_b$ and introduced the auxiliary coefficients
\bea
\label{co1}
b_0^{\pm}&=&u\pm1;\;\hspace{3mm}b_n^{\pm}=\pm2\gamma_c^{2n}(s)\hspace{3mm}\mathrm{if}\;n>0,\\
v_n^\pm&=&2n\pm u.
\eea
Carrying-out the integrals in Eq.~(\ref{s1}) with the expanded functions~(\ref{homt}), subtracting the bulk part of the grand potential, and switching to the dimensionless distance $\tzt=-\kappa_bz_t$ and polymer length $\tL=\kappa_bL$, one gets the polymer self-energy in the form
\be
\label{pst}
\Delta\Omega_{pp}(z_t)=\Delta\Omega_*\zeta(z_t),
\ee
with the characteristic energy $\Delta\Omega_*=\ell_B\tau^2/(2\kappa_b)$ and the dimensionless self-energy
\be\label{pst2}
\zeta(\tzt)=\int_1^\infty\frac{\mathrm{d}u}{u^2-1}\left\{2F(u)+\tdel(u)G^2(u)\right\}.
\ee
In Eq.~(\ref{pst2}), the delta function is given by Eq.~(\ref{del2}) and we introduced the auxiliary functions
\bea
\label{fnm}
F(u)&=&\sideset{}{'}\sum_{n,m\geq0}\frac{b_n^+b_m^-}{v_n^+v_m^-}e^{-(v_n^++v_m^-)\tzt}\\
&&\times\left\{1-e^{-v_n^+\tL}-\frac{v_n^+}{v_n^++v_m^-}\left[1-e^{-(v_n^++v_m^-)\tL}\right]\right\}\nonumber\\
\label{gnm}
G(u)&=&\sum_{n\geq0}\frac{b_n^+}{v_n^+}\left(1-e^{-v_n^+\tL}\right)e^{-v_n^+\tzt}.
\eea
In Eq.~(\ref{fnm}), the prime above the sum sign means that the term with indices $n=m=0$ corresponding to the bulk self-energy should not be included in the summation. In terms of the same dimensionless parameters, the MF grand potential~(\ref{pm5}) reads
\bea
\label{pm7}
\Omega_{pm}(\tzt)&=&\frac{2\tau}{q\kappa_b}\left\{\mathrm{Li}_2\left[\gamma_c(s)e^{-\tzt}\right]-\mathrm{Li}_2\left[-\gamma_c(s)e^{-\tzt}\right]\right.\nonumber\\
&&\left.-\mathrm{Li}_2\left[\gamma_c(s)e^{-\tzt-\tL}\right]+\mathrm{Li}_2\left[-\gamma_c(s)e^{-\tzt-\tL}\right]\right\}.\nonumber\\
\eea

\subsubsection{Thermodynamic limit $\tL\to\infty$}
\label{syel}
\begin{figure}
\includegraphics[width=1.2\linewidth]{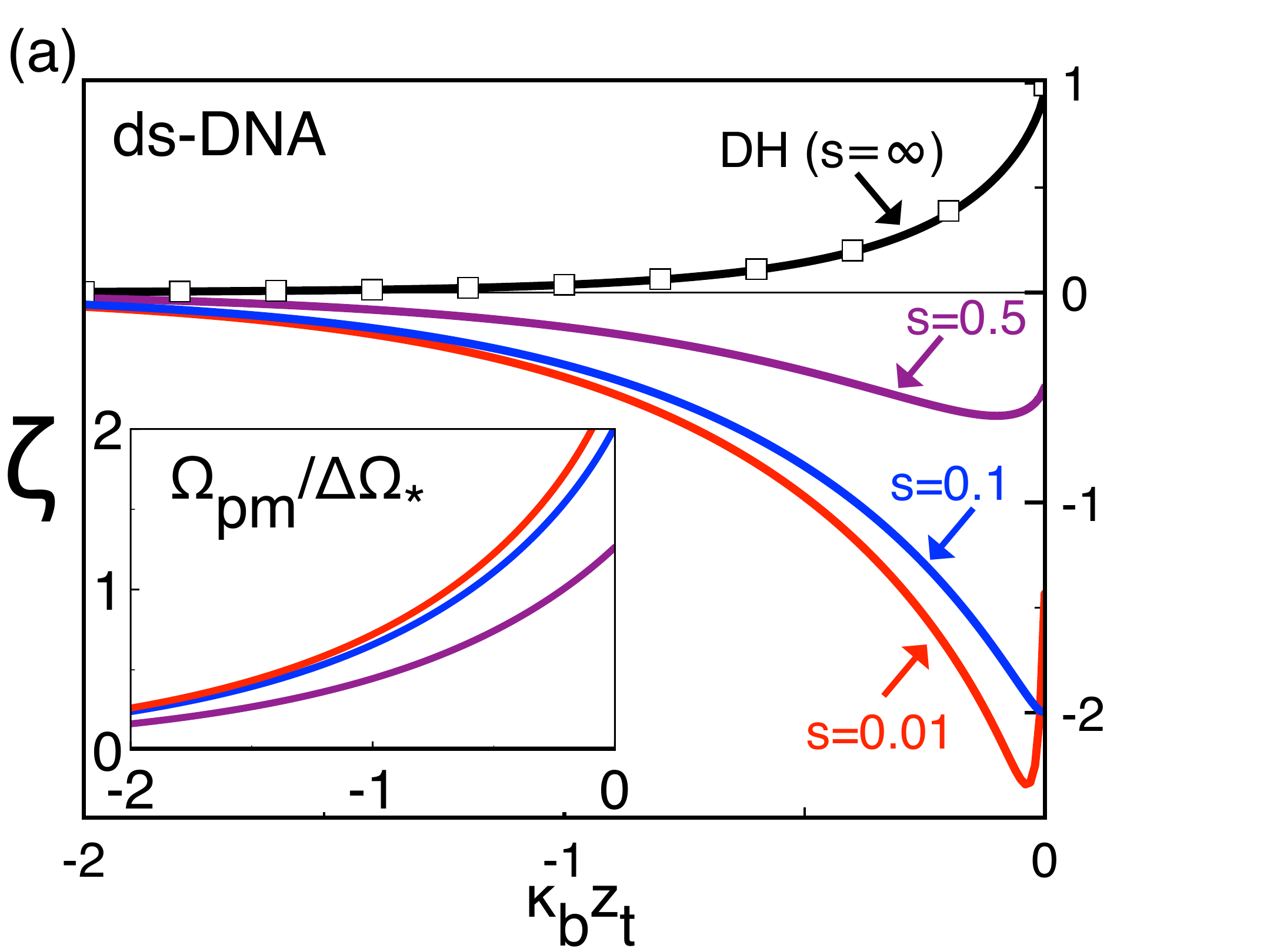}
\includegraphics[width=1.2\linewidth]{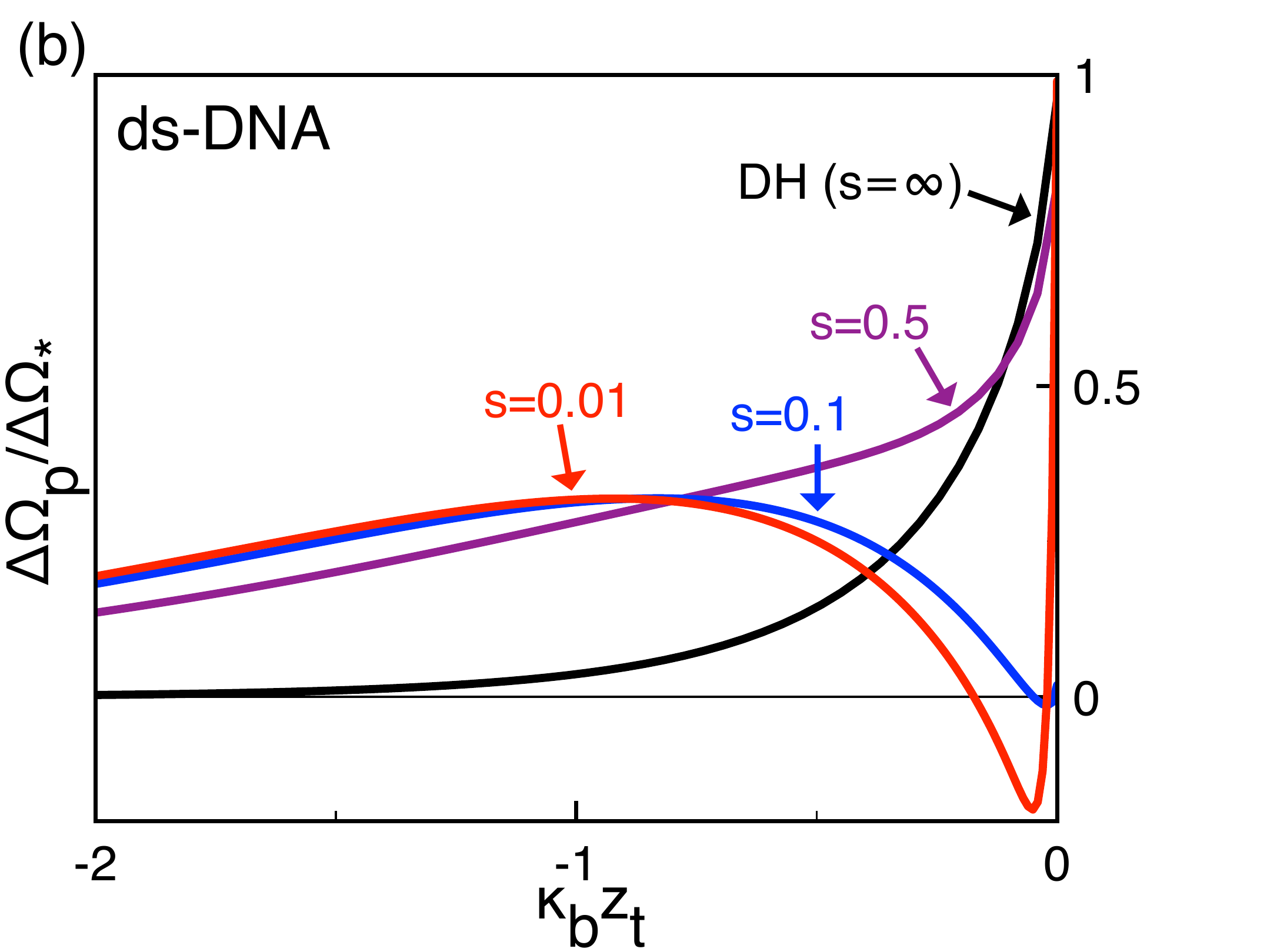}
\caption{(Color online) (a) Dimensionless self-energy~(\ref{pst2}) (main plot) and MF grand potential~(\ref{pm7}) rescaled by the characteristic energy $\Delta\Omega_*=\ell_B\tau^2/(2\kappa_b)$  (inset). (b) Rescaled total grand potential $\Delta\Omega_p=\Delta\Omega_{pp}+\Omega_{pm}$. The membrane permittivity is $\e_m=2$ and the ds-DNA charge density $\tau=0.59$ $e/${\AA}. The square symbols in (a) correspond to the DH limit Eq.~(\ref{pst3}).} 
\label{fig7}
\end{figure}

We consider first the regime of long ds-DNA molecules and take the thermodynamic limit $\tL\to\infty$. In this limit where the exponential terms of Eqs.~(\ref{fnm})-(\ref{pm7}) vanish, the total grand potential becomes independent of the polymer length. In Fig.~\ref{fig7}(a), we show that with the decrease the parameter $s$ from the DH limit ($s=\infty$) to the GC regime ($s=0.01$), the self-energy is transformed from repulsive to attractive (main plot) while the MF grand potential barrier rises (inset). Fig.~\ref{fig7}(b) shows that as a result of these effects, the grand potential becomes more repulsive outside the interfacial region $|z_t|\gtrsim\kappa_b$  but switches at $s\lesssim0.1$ from repulsive to attractive close to the surface. Hence, like-charge attraction is also expected for perpendicular ds-DNA molecules approaching strongly charged membranes. 

For an analytical insight into the behaviour of the self-energy, we fix the membrane permittivity to $\e_m=0$ and Taylor-expand Eq.~(\ref{pst2}) in the DH regime $s\gg1$ as $\zeta(\tzt)=\zeta^{(0)}_{DH}(\tzt)+s^{-2}\zeta^{(1)}_{DH}(\tzt)+O\left(s^{-4}\right)$.  Fig.~\ref{fig7}(a) displays by square symbols the DH contribution 
\be
\label{pst3}
\zeta^{(0)}_{DH}(\tzt)=e^{-2\tzt}+2\tzt\mathrm{Ei}(-2\tzt)
\ee
accounting for repulsive polymer-image charge interactions. The attractive correction term reads in turn
\bea
\label{pst4}
\zeta_{DH}^{(1)}(\tzt)&=&-\frac{1}{2}e^{2\tzt}\mathrm{Ei}(-6\tzt)+\left(1-\frac{1}{2}e^{2\tzt}\right)\mathrm{Ei}(-4\tzt)\nonumber\\
&&+\left(1+4\tzt^2-\frac{1}{2}e^{-2\tzt}\right)\mathrm{Ei}(-2\tzt)\nonumber\\
&&-\frac{1}{2}e^{-2\tzt}\left[2+\gamma-4\tzt+\ln\left(\frac{4\tzt}{3}\right)\right].
\eea
Based on these potentials, one finds that towards the membrane surface, the self-energy converges to
\be\label{pst5}
\zeta(0)=1-\frac{1}{s^2}+O\left(s^{-4}\right).
\ee
In Eq.~(\ref{pst5}), the negative sign of the beyond-DH correction characterizes the finite charge-induced reduction of the image-charge barrier (the first term) in Fig.~\ref{fig7}(a). Far away from the membrane surface $\tzt\gg1$, the self-energy is exponentially screened as
\be
\label{pst6}
\zeta(\tzt)\approx\frac{e^{-2\tzt}}{2\tzt}\left\{1-\frac{\tzt}{s^2}\left[\gamma+\ln\left(\frac{4\tzt}{3}\right)\right]\right\}\textcolor{black}{+O\left(s^{-4}\right)}.
\ee
In Eq.~(\ref{pst6}), the finite charge correction term is seen to be longer ranged than the DH contribution. Thus, similar to polymers parallel with the membrane, far enough from the membrane surface, solvation effects take over image charge forces and correlations bring an attractive contribution to the polymer grand potential. 

The blue curve in Fig.~\ref{fig7}(b) corresponds to the phase coexistence where the attractive well switches from the metastable to the stable state. We emphasize that in the present case of perpendicular polymers, the critical parameter value $s=0.1$ for phase coexistence is significantly lower than the value $s=0.5$ found for parallel polyelectrolytes.  In the case of polyethylene terephthalate membranes at $pH=7$, this value can be reached at the bulk salt density $\rho_b\approx0.02$ M. In order to better quantify the effect of the polymer orientation, in Fig.~\ref{fig5}, we plotted the critical line separating the parameter regimes with binding and unbinding polymers oriented perpendicular to the membrane surface (dashed curve). The comparison of the solid and dashed curves shows that at fixed polymer charge, the occurrence of the polymer attraction in the perpendicular configuration requires indeed 2-3 times larger membrane charges. One also notes that the critical line ends at the considerably larger polymer charge density $\tau_c=0.55$ $e/${\AA}. The weaker solvation effect for perpendicular polyelectrolytes stems from the fact that in this configuration, the polymer charges are only partially covered by the interfacial counterion cloud. 

In the final subsection we relax the thermodynamic limit and consider the finite length effects on polymer adsorption to like-charge membranes.
\begin{figure}
\includegraphics[width=1.2\linewidth]{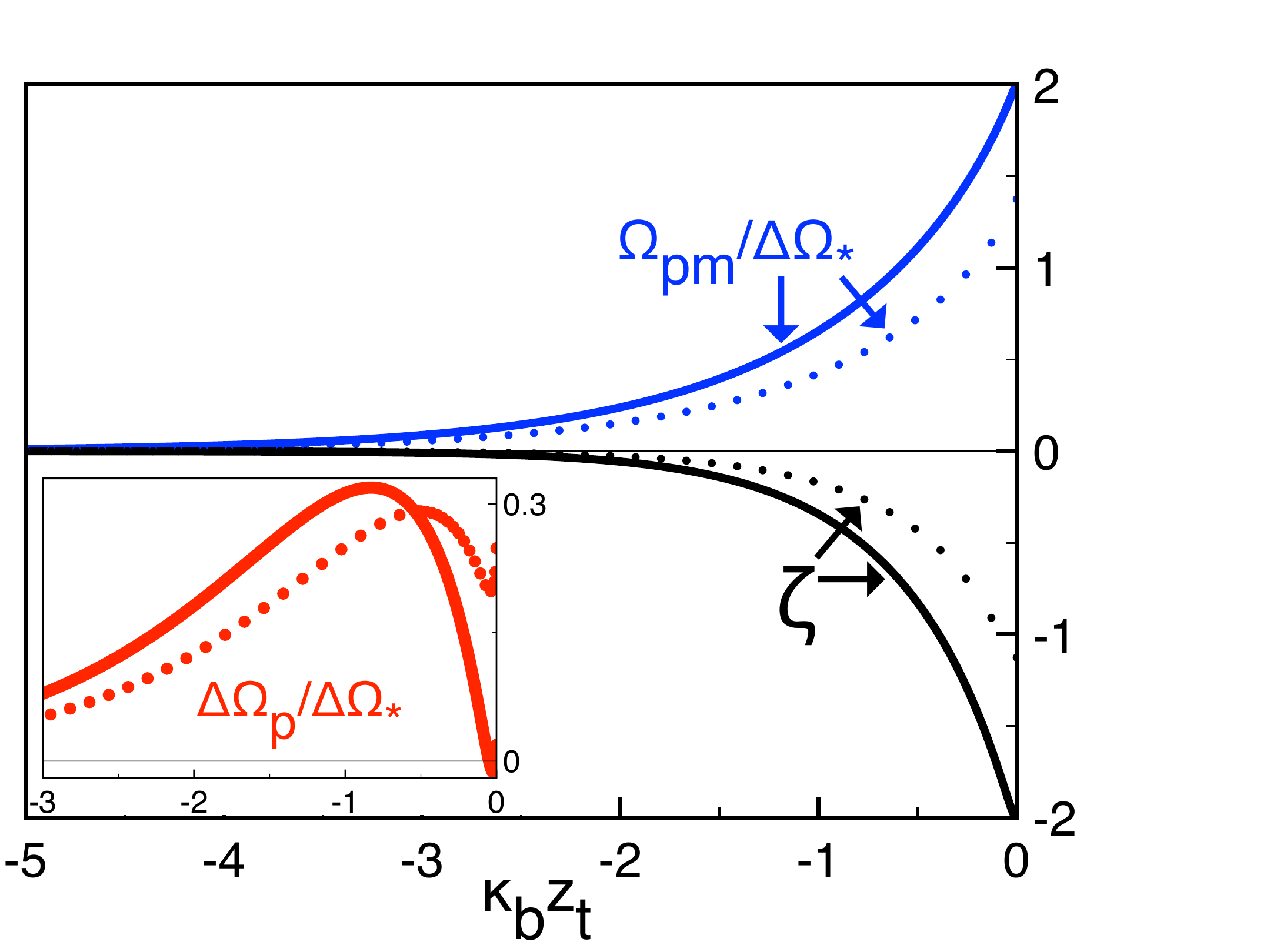}
\caption{(Color online) Main plot : MF grand potential~(\ref{pm7}) rescaled by the characteristic energy $\Delta\Omega_*=\ell_B\tau^2/(2\kappa_b)$ (blue) and dimensioneless self-energy~(\ref{pst2}) (black). Inset : Rescaled total grand potential $\Delta\Omega_p/\Delta\Omega_*$. Solid curves : $\tL=\infty$. Dotted curves : $\tL=1$. The membrane permittivity is $\e_m=2$, the ds-DNA charge density $\tau=0.59$ $e/${\AA}, and $s=0.1$.} 
\label{fig8}
\end{figure}

\subsubsection{Finite-size effects}

Fig.~\ref{fig8} compares the grand potential functions at the finite polymer length $\tL=1$ (dotted curves) and in the thermodynamic limit $\tL\to\infty$ (solid curves), at the coexistence value $s=0.1$ of Fig.~\ref{fig7}. One sees that the finiteness of the polymer length lowers the repulsive MF grand potential (blue curves) but rises the attractive self-energy (black curves). As a result, \textcolor{black}{with the reduction of the polymer length}, the one-loop grand potential (inset) becomes less repulsive far from the interface but also less attractive at the interface as the attractive minimum rises and becomes metastable. Hence, in the GC regime considered in Fig.~\ref{fig8}, the overall effect of the reduced polymer length is the attenuation of the like-charge attraction driven by the interfacial solvation force. Next, we characterize this finite length effect with analytical details by calculating the surface value of the self-energy~(\ref{pst2}) in the DH regime.
\begin{figure}
\includegraphics[width=1.05\linewidth]{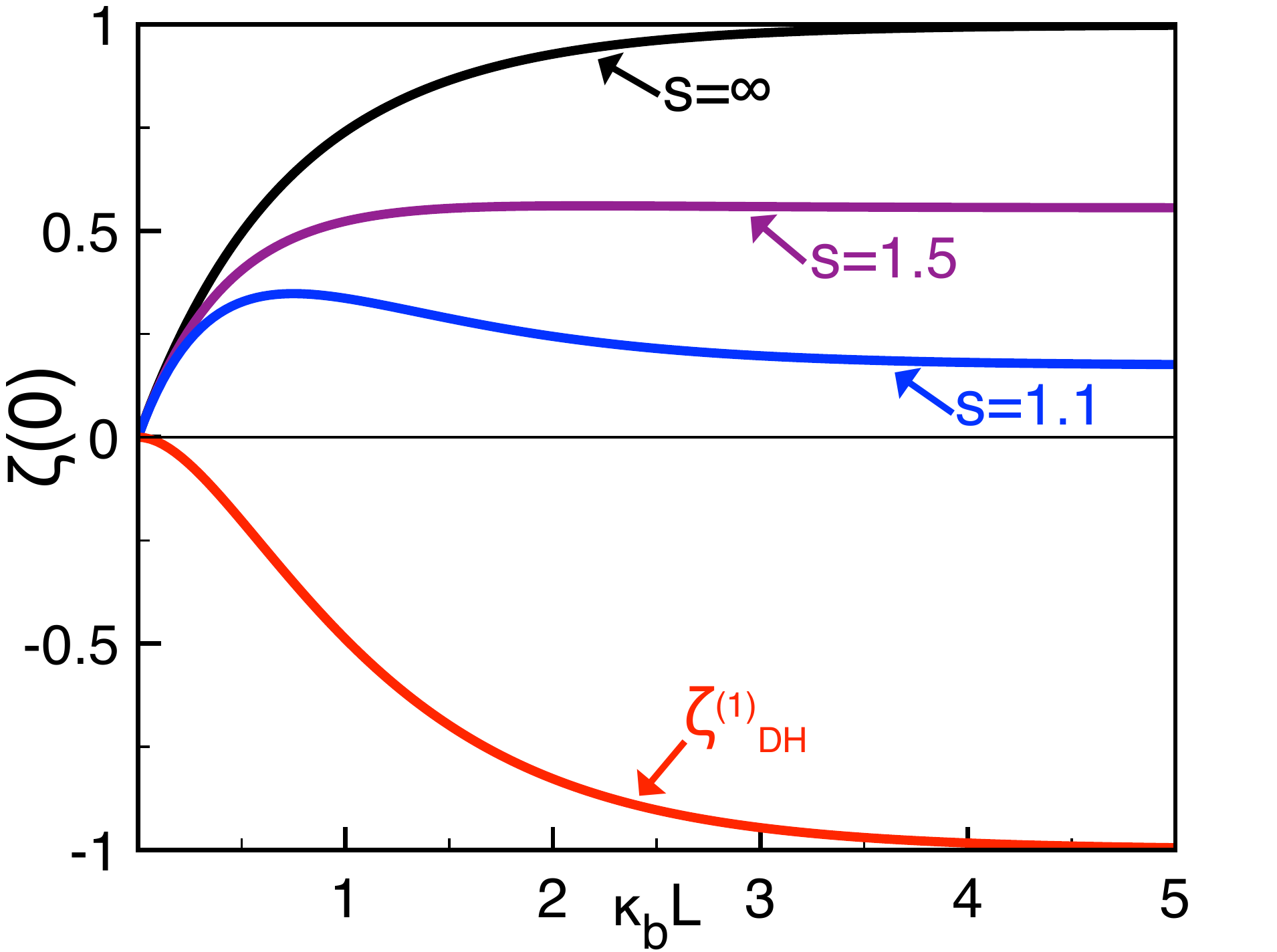}
\caption{(Color online) Surface self-energy $\zeta(0)=\zeta^{(0)}_{DH}(0)+s^{-2}\zeta^{(1)}_{DH}(0)$ (black, purple, and blue curves)  at various values of $s$ versus the polymer length $\tL$ from Eqs.~(\ref{pst7})-(\ref{pst8}). The red curve is the finite charge correction~(\ref{pst8}). The black curve at $s=\infty$ equally corresponds to the DH self-energy~(\ref{pst7}).} 
\label{fig9}
\end{figure}

The surface self-energy is relevant to translocation experiments since this quantity corresponds to the contribution from charge correlations to the work to be done in order to drive the polymer to the membrane surface. By Taylor-expanding Eq.~(\ref{pst2}) in the DH regime $s\gg1$, one gets $\zeta(0)=\zeta^{(0)}_{DH}(0)+s^{-2}\zeta^{(1)}_{DH}(0)$. The pure DH term is given by
\be
\label{pst7}
\zeta^{(0)}_{DH}(0)=\left(1-e^{-\tL}\right)^2+2\tL\left[\mathrm{Ei}(-2\tL)-\mathrm{Ei}(-\tL)\right],
\ee
and the \textcolor{black}{finite} membrane charge correction reads
\bea
\label{pst8}
\zeta^{(1)}_{DH}(0)&=&-\left(1-e^{-\tL}\right)^2+\left[2\tL-\frac{1}{2}\ln(4\tL)-\frac{\gamma}{2}\right]e^{-2\tL}\nonumber\\
&&-2\tL\;e^{-\tL}+\frac{1}{2}e^{-2\tL}\;\mathrm{Ei}(\tL)-\frac{1}{2}e^{2\tL}\;\mathrm{Ei}(-6\tL)\nonumber\\
&&-\left(\frac{1}{2}e^{2\tL}-1\right)\mathrm{Ei}(-4\tL)+\frac{1}{2}e^{2\tL}\;\mathrm{Ei}(-3\tL)\nonumber\\
&&+\left(4\tL^2+1-\frac{1}{2}e^{2\tL}\right)\mathrm{Ei}(-2\tL)\nonumber\\
&&-\left(2\tL^2+1\right)\mathrm{Ei}(-\tL).
\eea
Eqs.~(\ref{pst7}) and~(\ref{pst8}) are plotted in Fig.~\ref{fig9}. With increasing length, the repulsive DH potential~(\ref{pst7}) of positive value (black) and the attractive finite charge correction term~(\ref{pst8}) of negative value (red) are amplified until they saturate at polymer lengths $\tL\gtrsim2$. 

In Fig.~\ref{fig9}, we plotted the total self-energy  $\zeta(0)$ at various values of the parameter $s$ (black, purple, and blue curves). Based on the functions~(\ref{pst7}) and~(\ref{pst8}), one finds that for short polymers $\tL\ll1$, the self-energy rises algebraically with the polymer length as
\be
\label{pst9}
\zeta(0)\approx2\ln(2)\tL-\tL^2-\frac{2}{s^2}\ln(2)\tL^2\textcolor{black}{+O\left(s^{-4}\right)}.
\ee
For long polymers $\tL\gg1$, the self-energy converges exponentially to the thermodynamic limit of Eq.~(\ref{pst5}),
\be
\label{pst10}
\zeta(0)\approx1-\frac{2}{\tL}e^{-\tL}-\frac{1}{s^2}\left(1-\frac{16}{3\tL}e^{-\tL}\right)\textcolor{black}{+O\left(s^{-4}\right)},
\ee
which explains the saturation of the surface self-energy at the particularly low value $\tL\approx2$. We now note that below $s\approx2.0$ where one approaches the GC regime, the surface self-energy exhibits a peak and starts to drop beyond this point (blue curve). The location of the peak corresponds to the characteristic polymer length where the attractive interfacial solvation characterized by Eq.~(\ref{pst8}) takes over the image-charge barrier of Eq.~(\ref{pst7}). From the derivative of Eq.~(\ref{pst9}), an approximative value for the location of the bump follows as $\tL_*\approx s^2/2$.  Thus, the larger is the membrane charge (or the lower is the salt density), the shorter is the characteristic length $\tL_*$ (i.e. $\sigma_m\uparrow\tL_*\downarrow$ or $\rho_b\downarrow\tL_*\uparrow$). To summarize, in the DH regime $s\gtrsim2$, a larger polymer length means a more repulsive grand potential. Approaching the GC regime with $s\lesssim2.0$, the increase of the polymer length beyond the value $\tL_*$ results in turn in a less repulsive grand potential. This explains the finite-length behaviour of the grand potential in Fig.~\ref{fig8}. 

\begin{figure}
\includegraphics[width=1.05\linewidth]{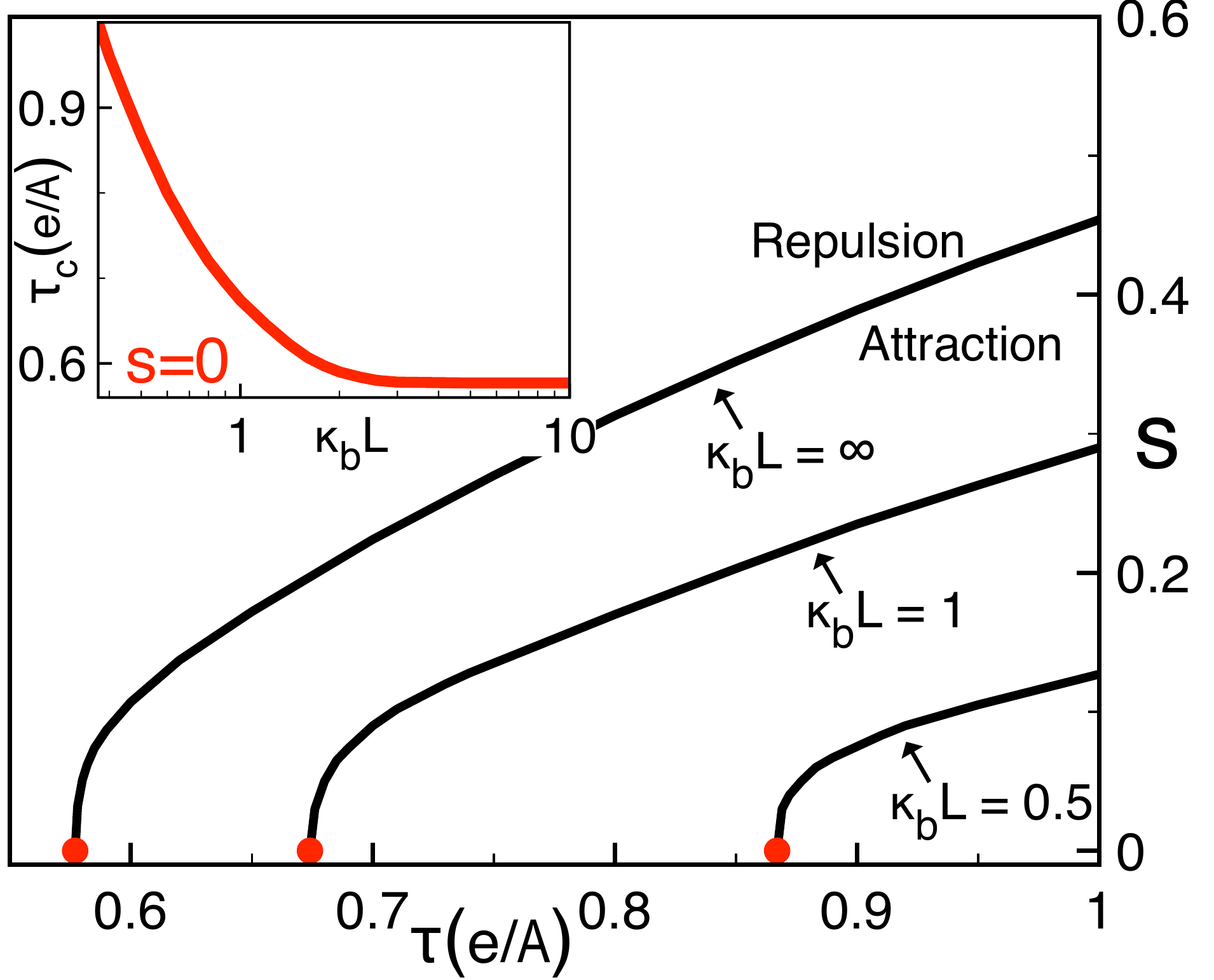}
\caption{(Color online) Phase diagram : Critical values of the parameter $s=\kappa_b\mu$ versus the polymer charge density $\tau$ separating the parameter regimes characterized by repulsion (area above the curves) and like-charge attraction (below the curves) at various polymer lengths given in the legend. The inset displays the polymer length dependence of the critical point where the critical lines of the main plot end in the GC limit ($s=0$).} 
\label{fig10}
\end{figure}

In order to highlight now the main finite size effects on the like-charge attraction, in Fig.~\ref{fig10}, we reported  the critical lines splitting the attraction and repulsion regimes in the thermodynamic limit $\tL\to\infty$, and at the finite polymer lengths $\tL=1$ and $\tL=0.5$. The phase diagram shows that at fixed polymer charge density $\tau$, the lower is the polymer length, the larger should be the membrane charge required to turn the membrane from repulsive to attractive (i.e. $L\downarrow\sigma_m\uparrow$). Furthermore, at fixed membrane charge (i.e. fixing $s$), the lower is the polymer length, the higher is the polymer charge required for the like-charge attraction to occur ($L\downarrow\tau\uparrow$). This is due to the fact that the weaker interfacial solvation for a shorter polymer has to be compensated by a higher polymer or membrane charge for the polymer-membrane interaction to remain attractive. In the inset of Fig.~\ref{fig10}, we plotted the length dependence of the critical points of the main plot (red dots) reached in the GC limit $s=0$. We note that at fixed polymer length (charge density), the red curve yields the lowest polyelectrolyte charge (length) where the like-charge attraction can occur. Increasing the polymer length, the critical polymer charge drops and saturates at $\tau_c\approx0.58$ $e/${\AA} for $\tL\gtrsim2$. This indicates that for ds-DNA molecules approaching strongly charged membranes in the perpendicular configuration, the similar charge attraction can be observed exclusively for sequence lengths larger than the characteristic value $L\simeq2/\kappa_b$. At the physiological salt concentration $\rho_b=0.1$ M, this lower bound corresponds to a molecule composed of $20$ bps. The predictions of the phase diagram in Fig.~\ref{fig10} calls for experimental verification.

\section{Conclusions}

In this work, we investigated electrostatic many-body effects on the interaction of polyelectrolytes with similarly-charged dielectric membranes in contact with an electrolyte solution. Our motivation stems from the fact that this configuration is frequently encountered in polymer transport experiments where both the DNA and the translocated membrane are negatively charged. Based on the one-loop expansion of the SC equations where the polymer was treated as a test charge, we calculated the polymer grand potential including charge correlations at the one-loop level. Our calculation generalizes previous DH-level formulations of polymer-membrane interactions~\cite{Netz2000,Buyuk2016} to strongly charged membranes.

We found that the polymer-membrane coupling is composed of three contributions. The direct coupling between the DNA and the membrane charges results in the classical MF-level similar charge repulsion. This is enhanced by the repulsive polymer-image charge interaction associated with the dielectric contrast between the solvent and the low permittivity membrane. The third effect is due to the deformation of the interfacial ionic cloud by the membrane charge, resulting in the enhanced interfacial screening of the polymer charges with respect to the bulk electrolyte. This \textcolor{black}{effect} lowers the polymer grand potential and translates into an attractive force oriented to the membrane surface. It is important to note that because \textcolor{black}{this peculiarity} originates from the non-uniform screening of the polymer charges, it could not be taken into account by the previous DH theories. In the present work, we considered both the parallel and the perpendicular polymer orientations, relevant to DNA adsorption and translocation experiments, respectively. In both cases, we found that for ds-DNA molecules at highly charged membranes, the attractive solvation interaction dominates the repulsive contributions and the molecule experiences a like-charge attraction to the membrane surface. Due to their lower charge density, this like-charge attraction does not occur for ss-DNA molecules. The physical conditions for the occurrence of the like-charge attraction are fully characterized by the phase diagrams of Figs.~\ref{fig5} and~\ref{fig10}.

The present formalism is based on \textcolor{black}{three} approximations. First, the polymer is modeled as a rigid line charge and \textcolor{black}{this simplified model neglects the contribution from the conformational fluctuations of the DNA molecule. The latter complication can be in principle taken into account by coupling the present theory with Edward's path integral formulation of fluctuating polymers. The corresponding MF equations of state were indeed derived in Ref.~\cite{dun} and the extension of this approach beyond-MF level was formally introduced. However, it should be noted that the application of this extended theory to the present inhomogeneous electrolyte system is a formidable task that still remains an open challenge.} Secondly, the polyelectrolyte being considered as a test charge, the theory does not account for the influence of the polymer on its ionic environment. The latter approximation was motivated by the fact that if one opts for the consideration of the polymer charges at the full one-loop level, one looses the plane symmetry. As a result, the analytical solution of the one-loop kernel equation~(\ref{forker}) becomes impossible. One could approximately overcome this limitation by introducing a numerical charge renormalisation procedure as in Refs.~\cite{netzvar,Buyuk2016II}. Because this will bring additional numerical complication and shadow the analytical transparency of the present work, we leave this improvement to a future work. \textcolor{black}{In addition, the present work focused exclusively on the parallel and perpendicular polymer configurations.  This constraint can be relaxed in the future by considering the rotational angle of the molecule as an internal degree of freedom of the system. Finally, it should be noted that the 1l-level evaluation of the polymer grand potential is expected to be valid up to the intermediate coupling regime. In the presence of high membrane charges and polyvalent ions that drive the system to the strong coupling regime, it would be appropriate to introduce strong-coupling corrections or semi-numerical approaches able to cover this regime~\cite{SC1,SC2,SC3,SC4}. The experimental observation of our prediction~\cite{Molina2014} indicates that despite these approximations, our theory can already capture the essential physics of like-charge adsorption. Moreover}, the numerous predictions of our work can be tested by \textcolor{black}{additional} experiments, or in simulations.

\smallskip
\appendix

\section{Calculating the polymer free energy} 
\label{ap1}

In this appendix, we calculate the electrostatic grand potential of a \textit{test polyelectrolyte} immersed in an electrolyte solution. Our starting point is the electrostatic variational grand potential of a charged liquid, derived in Ref.~\cite{netzvar} in the form
\bea\label{varf}
\Omega_v&=&-\frac{1}{2}\mathrm{Tr}\ln\left[v\right]+\int\mathrm{d}\br\sigma(\br)\psi(\br)\\
&&+\frac{k_BT}{2e^2}\int\mathrm{d}\br\e(\br)\left\{\nabla_\br\cdot\nabla_{\br'} \left.v(\br,\br')\right|_{\br'\to\br}-\left[\nabla\psi(\br)\right]^2\right\}\nonumber\\
&&-\sum_i\Lambda_i\int\mathrm{d}\br e^{-V_i(\br)}e^{-q_i\psi(\br)}e^{-\frac{q_i^2}{2}v(\br,\br)}.\nonumber
\eea
In Eq.~(\ref{varf}), the potential $v(\br,\br')$ stands for the variational Green's function, $\sigma(\br)$ is the density of fixed charges, $\psi(\br)$ the average electrostatic potential induced by the latter, $k_B$ the Boltzmann constant, $T$ the liquid temperature, and $e$ the electron charge. Furthermore, the function $\e(\br)$ corresponds to the dielectric permittivity profile, $\Lambda_i$ and $q_i$ are respectively the fugacity and the valency of an ion of species $i$. Finally, the function $V_i(\br)$ is the ionic steric potential accounting for the rigid boundaries in the system.

\subsection{Rescaling the electrostatic grand potential}

We consider now a symmetric electrolyte composed of two oppositely charged ionic species, each  with valency $q$ and bulk density $\rho_b$. The SC equations obtained from the extremization of the grand potential~(\ref{varf}) with respect to the average potential $\psi(\br)$ and Green's function $v(\br,\br')$ read~\cite{netzvar,Buyuk2010}
\bea\label{Eq1}
&&\nabla\e(\br)\cdot\nabla\psi(\br)-\frac{2\rho_bqe^2}{k_BT}e^{-V_i(\br)-\frac{q^2}{2}\delta v(\br)}\sinh\left[q\psi(\br)\right]\nonumber\\
&&=-\frac{e^2}{k_BT}\sigma(\br)\\
\label{Eq2}
&&\nabla\e(\br)\cdot\nabla v(\br,\br')\nonumber\\
&&-\frac{2\rho_bq^2e^2}{k_BT}e^{-V_i(\br )-\frac{q^2}{2}\delta v(\br)}\cosh\left[q\psi(\br)\right]v(\br,\br')\nonumber\\
&&=-\frac{e^2}{k_BT}\delta(\br-\br'),
\eea
with the ionic self-energy defined as
\be\label{Eq3}
\delta v(\br)=\lim_{\br'\to\br}\left\{v(\br,\br')-v_b(\br-\br')\right\}.
\ee
In Eq.~(\ref{Eq3}), we introduced the bulk Debye-H\"{u}ckel potential $v_b(\br-\br')=\ell_Be^{-\kappa_b|\br-\br'|}/|\br-\br'|$, with the DH screening parameter $\kappa_b^2=8\pi q^2\ell_B\rho_b$ and the Bjerrum length $\ell_B=e^2/(4\pi\e_wk_BT)$ where $\e_w=80$ is the dielectric permittivity of water. Furthermore, we used  the relation $\rho_b=\Lambda_i\;e^{-\frac{q_i^2}{2}v_b(0)}$ between the ionic fugacity and bulk density. 

We consider now a simple symmetric electrolyte without DNA. The electrolyte is in contact with a charged plane of infinite thickness, with the interface located at $z=0$ and carrying the smeared charge distribution $\sigma(z)=-\sigma_s\delta(z)$. We also assume that the membrane has the same permittivity as the solvent, i.e. $\e(\br)=\e_w$. For the one-loop (1l) expansion that will be carried out next, we introduce the new average potential $\phi(z)=q\psi(z)$ and Green's function $u(\br)=q^2v(\br,\br')$. Defining as well the adimensional parameter $s=\kappa_b\mu$ and the electrostatic coupling parameter $\Gamma=q^2\kappa_b\ell_B$~\cite{Buyuk2012}, rescaling all lengths according to $\tbr=\kappa\br$, and using Eq.~(\ref{Eq3}), the variational grand potential~(\ref{varf}) takes the adimensional form
\bea\label{Eq5}
\Omega_v&=&-\frac{1}{2}\mathrm{Tr}\ln\left[u/q^2\right]-\frac{1}{2\pi s\Gamma}\int\mathrm{d}\tbr\delta(\tz)\phi(\tbr)\\
&&+\frac{1}{8\pi\Gamma}\int\mathrm{d}\tbr\left\{\left.\nabla_{\tbr}\cdot\nabla_{\tbr'}u(\tbr,\tbr')\right|_{\tbr\to\tbr'}-\left[\nabla\phi(\tbr)\right]^2\right\}\nonumber\\
&&-\frac{1}{4\pi\Gamma}\int\mathrm{d}\tbr\cosh\left[\phi(\tbr)\right]e^{-\frac{1}{2}\delta u(\tbr)}.\nonumber
\eea
In the same dimensionless variables, the SC Eqs.~(\ref{Eq1})-(\ref{Eq2}) read in turn
\bea
\label{Eq6}
&&\nabla^2\phi(\tbr)-\sinh\left[\phi(\tbr)\right]e^{-\frac{1}{2}\delta u(\tbr)}=\frac{2}{s}\delta(\tz)\\
\label{Eq7}
&&\nabla^2u(\tbr,\tbr')-\cosh\left[\phi(\tbr)\right]e^{-\frac{1}{2}\delta u(\tbr)}u(\tbr,\tbr')=-4\pi\Gamma\delta(\tbr-\tbr').\nonumber\\
\eea
One can verify that Eqs.~(\ref{Eq6})-(\ref{Eq7}) follow directly from the extremization of the grand potential~(\ref{Eq5}) with respect to the rescaled potentials $\phi(\tbr)$ and $u(\tbr,\tbr')$.

\subsection{1l expansion of the SC equations}

We will now expand the rescaled grand potential~(\ref{Eq5}) and the SC Eqs.~(\ref{Eq6})-(\ref{Eq7}) at 1l-order. This corresponds to a Taylor-expansion of these equations at the linear order in the coupling parameter $\Gamma$~\cite{Buyuk2012}. \textcolor{black}{Eq.~(\ref{Eq7}) shows} that at leading order,  the propagator $u(\tbr,\tbr')$ is proportional to the electrostatic coupling parameter $\Gamma$. Thus, we expand the electrostatic potentials as
\bea
\label{Eq8}
&&\phi(\tbr)=\phi_0(\tbr)+\Gamma\phi_1(\tbr)+O\left(\Gamma^2\right)\\
\label{Eq9}
&&u(\tbr,\tbr')=\Gamma u_1(\tbr,\tbr')+O\left(\Gamma^2\right).
\eea
Inserting Eqs.~(\ref{Eq8})-(\ref{Eq9}) into the SC Eqs.~(\ref{Eq6})-(\ref{Eq7}) and Taylor-expanding the latter at the order $O\left(\Gamma\right)$, one finds
\bea
\label{Eq10}
&&\nabla^2\phi_0(\tbr)-\sinh\left[\phi_0(\tbr)\right]=\frac{2}{s}\delta(\tz)\\
\label{Eq11}
&&\nabla^2\phi_1(\tbr)-\cosh\left[\phi_0(\tbr)\right]\phi_1(\tbr)=-\frac{1}{2}\sinh\left[\phi_0(\tbr)\right]\delta u_1(\tbr)\nonumber\\
&&\\
\label{Eq12}
&&\nabla^2u_1(\tbr,\tbr')-\cosh\left[\phi_0(\tbr)\right]u_1(\tbr,\tbr')=-4\pi\delta(\tbr-\tbr').\nonumber\\
\eea
Eq.~(\ref{Eq10}) is the MF-level equation of state, i.e. the PB equation for the MF potential $\phi_0(\tbr)$. The solution of Eq.~(\ref{Eq11}) $\phi_1(\tbr)$ yields in turn the 1l-level correlation corrections to the MF average potential. Finally, the solution of Eq.~(\ref{Eq12}) corresponds to the 1l-level electrostatic propagator accounting for ionic correlations. 

Inserting now the expanded potentials~(\ref{Eq8}) and~(\ref{Eq9}) into the variational grand potential~(\ref{Eq5}), expanding the latter in the coupling parameter $\Gamma$ up to the order $O(\Gamma)$, the 1l-level grand potential follows as
\be\label{1llev}
\Omega_{1l}=\frac{1}{\Gamma}\Omega_{MF}+\Omega_u+\Gamma\Omega_{\phi_1}.
\ee
In Eq.~(\ref{1llev}), the rescaled MF grand potential reads
\bea\label{Eq13}
\Omega_{MF}&=&-\frac{1}{2\pi s}\int\mathrm{d}\tbr\delta(\tz)\phi_0(\tbr)-\frac{1}{8\pi}\int\mathrm{d}\tbr\left[\nabla\phi_0(\tbr)\right]^2\nonumber\\
&&-\frac{1}{4\pi}\int\mathrm{d}\tbr\cosh\left[\phi_0(\tbr)\right], 
\eea
and the correction terms associated with the Green's function and the average potential correction are 
\bea\label{Eq14}
\Omega_u&=&-\frac{1}{2}\mathrm{Tr}\ln\left[u/q^2\right]\\
&&+\frac{1}{8\pi}\int\mathrm{d}\tbr\mathrm{d}\tbr'\delta(\tbr-\tbr')\nabla_{\tbr}\cdot\nabla_{\tbr'}u_1(\tbr,\tbr')\nonumber\\
&&+\frac{1}{8\pi}\int\mathrm{d}\tbr\cosh\left[\phi_0(\tbr)\right]\delta u_1(\tbr)\nonumber\\
\label{Eq15}
\Omega_{\phi_1}&=&-\frac{1}{8\pi}\int\mathrm{d}\tbr\left[\nabla\phi_1(\tbr)\right]^2\\
&&-\frac{1}{8\pi}\int\mathrm{d}\tbr\left\{\cosh\left[\phi_0(\tbr)\right]\phi_1^2(\tbr)\right.\nonumber\\
&&\left.\hspace{1.7cm}-\sinh\left[\phi_0(\tbr)\right]\delta u_1(\tbr)\phi_1(\tbr)\right\}.\nonumber
\eea
One can verify that the extremization of the functionals~(\ref{Eq13}), (\ref{Eq14}), and~(\ref{Eq15}) with respect to the potentials $\phi_0(\tbr)$, $u_1(\tbr,\tbr')$, and $\phi_1(\tbr)$, respectively, yields the equations~(\ref{Eq10})-(\ref{Eq12}) solved by these potentials.

\subsection{Computing the polymer free energy}

In the present work, we will compute the polymer grand potential by restricting ourselves to the lowest order contribution~(\ref{Eq13}) to Eq.~(\ref{1llev}). To this aim, we restore the physical parameters via the inverse transformations $\tbr\to\br=\tbr/\kappa_b$ and $\psi_0(\br)=\phi_0(\tbr)/q$. The MF-level grand potential~(\ref{Eq13}) reads
\bea\label{Eq16}
\Omega_{MF}&=&-\frac{k_BT}{2e^2}\int\mathrm{d}\br\e(\br)\left[\nabla\psi_0(\br)\right]^2+\int\mathrm{d}\br\sigma(\br)\psi_0(\br)\nonumber\\
&&-2\rho_b\int\mathrm{d}\br e^{-V_i(\br)}\cosh\left[q\psi_0(\br)\right].
\eea
The MF-level equation of state~(\ref{Eq10}) and the kernel Eq.~(\ref{Eq12}) are 
\bea\label{Eq17}
&&\nabla\e(\br)\cdot\nabla\psi_0(\br)-\frac{2\rho_bqe^2}{k_BT}e^{-V_i(\br)}\sinh\left[q\psi_0(\br)\right]\nonumber\\
&&=-\frac{e^2}{k_BT}\sigma(\br)\\
\label{Eq18}
&&\nabla\e(\br)\cdot\nabla v(\br,\br')-\frac{2\rho_bq^2e^2}{k_BT}e^{-V_i(\br )}\cosh\left[q\psi_0(\br)\right]v(\br,\br')\nonumber\\
&&=-\frac{e^2}{k_BT}\delta(\br-\br').
\eea
For the calculation that follows, the 1l correction to the PB Eq.~(\ref{Eq15}) will not be needed.

We include now the charged polymer located close to the single charged membrane (see Fig.~\ref{fig1}). The total charge density is composed of the polymer and the membrane charges,
\be
\label{Eq19}
\sigma(\br)=\sigma_m(\br)+\sigma_p(\br).
\ee
In the following derivation of the polymer grand potential, the potential induced by the membrane charge $\sigma_m(\br)$ will be considered at the full non-linear level while the potential associated with the polymer charge $\sigma_p(\br)$ will be taken into account at the linear level (i.e. DH level). The theory will be thus valid for weakly charged polymers but there is no restriction on the strength of the membrane charge. Next, based on the superposition principle, we express the total potential as the sum of the polymer and membrane charge contributions,
\be
\label{Eq20}
\psi_0(\br)=\psi_{0m}(\br)+\psi_{0p}(\br).
\ee
Inserting the decomposition~(\ref{Eq20}) into Eqs.~(\ref{Eq17})-(\ref{Eq18}) and expanding them at the linear order in the polymer potential $\psi_p(\br)$, one finds
\bea\label{Eq21}
&&\nabla\e(\br)\cdot\nabla\psi_{0m}(\br)-\frac{2\rho_bqe^2}{k_BT}e^{-V_i(\br)}\sinh\left[q\psi_{0m}(\br)\right]\nonumber\\
&&=-\frac{e^2}{k_BT}\sigma_m(\br)\\
\label{Eq22}
&&\left\{\nabla\e(\br)\cdot\nabla-\frac{2\rho_bq^2e^2}{k_BT}e^{-V_i(\br)}\cosh\left[q\psi_{0m}(\br)\right]\right\}\psi_{0p}(\br)\nonumber\\
&&=-\frac{e^2}{k_BT}\sigma_p(\br)\\
\label{Eq23}
&&\nabla\e(\br)\cdot\nabla v(\br,\br')-\frac{2\rho_bq^2e^2}{k_BT}e^{-V_i(\br )}\cosh\left[q\psi_{0m}(\br)\right]v(\br,\br')\nonumber\\
&&=-\frac{e^2}{k_BT}\delta(\br-\br').
\eea
The need for this decomposition will become clear below. One notes that Eq.~(\ref{Eq21}) is the non-linear PB equation for the potential $\psi_{0m}(\br)$ induced exclusively by the membrane charge. Eq.~(\ref{Eq22}) is in turn the linearized PB equation for the potential $\psi_{0p}(\br)$ associated with the polymer charge. The essential point is that in Eq.~(\ref{Eq22}), the screening term behind the polymer potential is non-uniform. Indeed, the latter corresponds to the local screening of the polymer potential in the ionic environment shaped by ion-membrane charge interactions. This is the point where the kernel Eq.~(\ref{Eq23}) becomes useful. By defining the kernel operator associated with Eq.~(\ref{Eq23})
\bea\label{Eq24}
v^{-1}(\br,\br')&=&\left[-\frac{k_BT}{e^2}\nabla\e(\br)\cdot\nabla\right.\\
&&\hspace{3mm}\left.+2\rho_bq^2\cosh\left[q\psi_{0m}(\br)\right]\right]\delta(\br-\br'),\nonumber
\eea
one can express Eq.~(\ref{Eq22}) as
\be
\label{Eq25}
\int\mathrm{d}\br'v^{-1}(\br,\br')\psi_{0p}(\br')=\sigma_p(\br).
\ee
By using the definition of the Green's function
\be
\label{Eq26}
\int\mathrm{d}\br''v^{-1}(\br,\br'')v(\br'',\br')=\delta(\br-\br'),
\ee
one can invert Eq.~(\ref{Eq25}) and express the potential induced by the polymer charge in terms of the Green's function solving Eq.~(\ref{Eq23}),
\be\label{Eq27}
\psi_{0p}(\br)=\int\mathrm{d}\br'v(\br,\br')\sigma_p(\br').
\ee

We insert now the decomposition~(\ref{Eq20}) into the grand potential~(\ref{Eq16}) and expand the latter in the polymer potential $\psi_{0p}(\br)$. By using as well the PB equation~(\ref{Eq21}) for the membrane potential, after some algebra, one finds
\be
\label{Eq28}
\Omega_{MF}=\Omega_{m}+\Omega_{p},
\ee
with the membrane and polymer \textcolor{black}{grand} potentials
\bea\label{Eq29}
\Omega_{m}&=&-\frac{k_BT}{2e^2}\int\mathrm{d}\br\e(\br)\left[\nabla\psi_{0m}(\br)\right]^2+\int\mathrm{d}\br\sigma_m(\br)\psi_{0m}(\br)\nonumber\\
&&-2\rho_b\int\mathrm{d}\br e^{-V_i(\br)}\cosh\left[q\psi_{0m}(\br)\right]\\
\label{Eq30}
\Omega_{p}&=&-\frac{k_BT}{2e^2}\int\mathrm{d}\br\e(\br)\left[\nabla\psi_{0p}(\br)\right]^2\nonumber\\
&&-\rho_bq^2\int\mathrm{d}\br e^{-V_i(\br)}\cosh\left[q\psi_{0m}(\br)\right]\psi_{0p}^2(\br)\nonumber\\
&&+\int\mathrm{d}\br\sigma_p(\br)\left[\psi_{0m}(\br)+\psi_{0p}(\br)\right].
\eea
By using the kernel~(\ref{Eq24}), one can express Eq.~(\ref{Eq30}) as
\bea
\label{Eq31}
\Omega_p&=&-\frac{1}{2}\int\mathrm{d}\br\mathrm{d}\br'\psi_{0p}(\br)v^{-1}(\br,\br')\psi_{0p}(\br')\nonumber\\
&&+\int\mathrm{d}\br\sigma_p(\br)\left[\psi_{0p}(\br)+\psi_{0m}(\br)\right].
\eea
Inserting into Eq.~(\ref{Eq31}) the expression~(\ref{Eq27}) for the electrostatic potential induced by the polymer charge, the polymer grand potential  finally takes the form
\be
\label{Eq32}
\Omega_p=\frac{1}{2}\int\mathrm{d}\br\mathrm{d}\br'\sigma_p(\br)v(\br,\br')\sigma_p(\br')+\int\mathrm{d}\br\sigma_p(\br)\psi_{0m}(\br).
\ee
Eq.~(\ref{Eq32}) indicates that the evaluation of the polymer grand potential necessitates the solution of the PB equation~(\ref{Eq21}) associated with the membrane charge and the 1l-level kernel equation~(\ref{Eq23}). 
\\

\textcolor{black}{{\bf Acknowledgement.} We thank Dr. Molina for bringing Ref. \cite{Molina2014} to our attention. This work is supported in part by the ANR grant `FSCF',
ANR-12-BSV5-0009-03.}


\begin{thebibliography}{99}
\bibitem {prc} Obtaining Eq.~(\ref{pm2}), we subtracted a constant bulk contribution diverging logarithmically in the counterion-only limit and rescaled the polymer-membrane grand potential by its value on the surface. We emphasize that adding a constant to the grand potential does not change the physics of the model. Indeed, the physically relevant function is the grand potential difference between the bulk and the location of the polymer, and the added constant does not play a role in the potential difference.
\end{thebibliography}
\end{document}